\documentclass[fleqn]{jfm}

\ifCUPmtlplainloaded \else
  \checkfont{eurm10}
  \iffontfound
    \IfFileExists{upmath.sty}
      {\typeout{^^JFound AMS Euler Roman fonts on the system,
                   using the 'upmath' package.^^J}%
       \usepackage{upmath}}
      {\typeout{^^JFound AMS Euler Roman fonts on the system, but you
                   dont seem to have the}%
       \typeout{'upmath' package installed. JFM.cls can take advantage
                 of these fonts,^^Jif you use 'upmath' package.^^J}%
      }
  \else
  \fi
\fi

\ifCUPmtlplainloaded \else
  \checkfont{msam10}
  \iffontfound
    \IfFileExists{amssymb.sty}
      {\typeout{^^JFound AMS Symbol fonts on the system, using the
                'amssymb' package.^^J}%
       \usepackage{amssymb}%
       \let\le=\leqslant  \let\leq=\leqslant
       \let\ge=\geqslant  
      }{}
  \fi
\fi

\ifCUPmtlplainloaded \else
  \IfFileExists{amsbsy.sty}
    {\typeout{^^JFound the 'amsbsy' package on the system, using it.^^J}%
     \usepackage{amsbsy}}
    {}
\fi

\newsavebox{\astrutbox}
\sbox{\astrutbox}{\rule[-5pt]{0pt}{20pt}}

\newcommand\etal{\mbox{\textit{et al.}}}

\newcounter{saveeqn}
\usepackage{epsfig}
\usepackage{psfrag}
\usepackage{amsfonts}          
\usepackage{amssymb}           
\usepackage{amsmath}           

\usepackage{hhline}            

\renewcommand{\thesection}{\arabic{section}}

\newcounter{eee}

\newcounter{eeeb}

\newcommand{\mypsfrag}[2]{\psfrag{#1}{\footnotesize{#2}}}
\newcommand{\npsfrag}[3]{\psfrag{#1}[#2]{\footnotesize{#3}}}

\newcommand{\x}{\times}

\newcommand{\dd}{\partial}
\newcommand{\de}{{\rm \, d}}

\renewcommand{\vec}[1]{\mbox{\boldmath $ #1$}}
\renewcommand{\v}{\vec}
\newcommand{\bec}{\vec}

\renewcommand{\P}{$P\ $}

\newcommand\ie{i.e.\ }


\title[Prandtl number dependence of convective dynamos]{Prandtl number dependence of
  convection driven dynamos in rotating spherical fluid shells}

\author[R. Simitev and F. H. Busse]%
{R.\ns S\ls I\ls M\ls I\ls T\ls E\ls V\ns
\and  F.\ns H.\ns B\ls U\ls S\ls S\ls E\ns}

\affiliation{Institute of Physics, University of Bayreuth, D-95440
  Bayreuth, Germany\\
{\tt radostin.simitev@uni-bayreuth.de}} 

\pubyear{2005}
\volume{532}
\pagerange{365--388}
\doi{S0022112005004398}
\date{1 June 2004 and in revised form 20 Jan 2005}
\begin{document}

\maketitle

\begin{abstract}
The value of the Prandtl number $P$ exerts a strong influence on
convection driven dynamos in rotating spherical shells filled with
electrically conducting fluids. Low Prandtl numbers promote dynamo
action through the shear provided by differential rotation, while the
generation of magnetic fields is more difficult to sustain in high
Prandtl number fluids where higher values of the magnetic Prandtl
number $P_m$ are required. The magnetostrophic approximation often used in
dynamo theory appears to be valid only for relatively high values of
$P$ and $P_m$. Dynamos with a minimum value of $P_m$
seem to be most readily realizable in the presence of convection
columns at moderately low values of $P$. The structure of the magnetic
field varies strongly with $P$ in that dynamos with a strong axial
dipole field are found for high values of $P$ while the energy of this
component is exceeded by that of the axisymmetric toroidal field and by
that of the non-axisymmetric components at low values of $P$. Some
conclusions are discussed in relation to the problem of the generation
of planetary magnetic fields by motions in their electrically
conducting liquid cores.  
\end{abstract}

\section{Introduction}

The problem of the generation of magnetic fields by motions of an
electrically conducting fluid in rotating spherical shells is one of
the fundamental problems of planetary and astrophysical sciences. The
increasing availability in recent years of computer capacity has
permitted large scale numerical simulations of this process, but the
creation of realistic models for planetary and stellar dynamos has
been hampered by a lack of knowledge about appropriate external
parameters. Since molecular values of material properties are usually
not attainable in computer simulations because of the limited
numerical resolution, eddy diffusivities must be invoked for
comparisons with observations. Eddy diffusivities represent the
effects of the unresolved scales of the turbulent velocity field and
it is often assumed for this reason that the eddy diffusivities for
velocities, temperature and magnetic fields are identical. 
The effects of turbulence on the diffusion of vector and scalar
quantities differ, however, and large differences in the 
corresponding molecular diffusivities are likely to be reflected in
the effective diffusivities caused by the fluctuating velocity 
field of the unresolved scales. In the case of the Earth's core, for
example, the magnetic diffusivity is assumed to have a value of the
order of 2 $m^2/sec$ (Braginsky and Roberts, 1995) which exceeds the
kinematic viscosity by a factor of the order $10^6$. Numerical
simulations are thus capable of resolving magnetic fields, but are 
far from resolving velocity structures. However, since the largest
unresolved scales, $v$, $l$, of velocity and length yield values for an
eddy viscosity $\nu_e \approx v\cdot l$ of the order 1 $m^2/sec$ or
less, it is reasonable to assume magnetic Prandtl numbers less than
unity. Since the concept of eddy diffusivities is simplistic there
have not been many theoretical derivations for ratios of turbulent
diffusivities which  enter the dimensionless equations for the
numerical simulations. Some theoretical considerations in the
astrophysical context can be found in the paper by Eschrich and
R\"udiger (1983). Values for an effective Prandtl number can
eventually be derived from experiments on turbulent 
convection (see, for example, Ahlers and Xu, 2001). Additional
complications arise in planetary and astrophysical applications
through anisotropies introduced by the effects of rotation and the
presence of large scale magnetic fields. While anisotropic eddy
diffusivity tensors will be useful for more realistic models, for the
purpose of the present paper we prefer the simplicity of scalar
diffusivities.  

After introducing the basic equations and the method of their
numerical solution in section 2 we shall consider the most important
properties of convection without magnetic field as a function of the
Prandtl number in section 3. Since a number of earlier papers on this
topic have been published this section can be kept relatively
short. In section 4 the onset of convection driven dynamos in fluids
with different Prandtl numbers is described. Energy aspects are
considered in section 5 and the validity of the magnetostrophic
approximation is  discussed in section 6. The influences of various
boundary conditions are considered in section 7 and a concluding
discussion is given in the final section 8. 
\begin{figure}
\vspace*{4mm}
\mypsfrag{z}{$z$}
\mypsfrag{x}{$x$}
\mypsfrag{y}{$y$}
\mypsfrag{r}{$\v r$}
\mypsfrag{th}{$\theta$}
\mypsfrag{phi}{$\varphi$}
\mypsfrag{om}{$\v \Omega$}
\mypsfrag{d}{$d$}
\begin{center}
\hspace*{10mm}
\epsfig{file=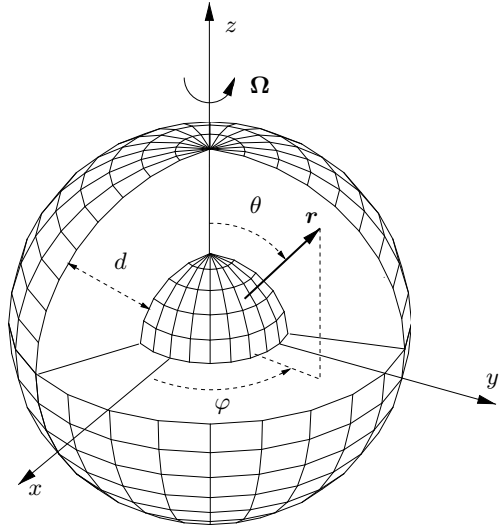,height=7cm,clip=}
\end{center}
\caption[]{Geometrical configuration of the problem. A part of the
 outer spherical surface is removed to expose the interior of the
 shell to which the conducting fluid is confined.}
\label{f.01}
\end{figure}

\section{Mathematical formulation of the problem and methods of solution}
We consider a rotating spherical fluid shell as shown in figure
\ref{f.01}. We assume that a 
static state exists with the temperature distribution $T_S = T_0 - \beta
d^2 r^2 /2$. Here $rd$ is the length of
the position vector with respect to the center of the sphere.
The gravity field is given by $\vec g = - d \gamma \vec r$. In addition to  $d$, the
time $d^2 / \nu$,  the temperature $\nu^2 / \gamma \alpha d^4$ and 
the magnetic flux density $\nu ( \mu \varrho )^{1/2} /d$ are used as
scales for the dimensionless description of the problem  where $\nu$ denotes
the kinematic viscosity of the fluid, $\kappa$ its thermal diffusivity,
$\varrho$ its density and $\mu$ is its magnetic permeability.
The equations of motion for the velocity vector $\vec u$, the heat
equation for the deviation 
$\Theta$ from the static temperature distribution, and the equation of
induction for the magnetic flux density $\vec B$ are thus given by 
\begin{subequations}
\begin{align}
\label{1a}
& 
\partial_t \vec{u} + \vec u \cdot \nabla \vec u + \tau \vec k \times
\vec u = - \nabla \pi +\Theta \vec r + \nabla^2 \vec u + \vec B \cdot
\nabla \vec B, \\
\label{1b}
&
\nabla \cdot \vec u = 0, \\
\label{1c}
&P(\partial_t \Theta + \vec u \cdot \nabla \Theta) = R \vec r \cdot \vec u + \nabla^2 \Theta, \\
&
\nabla \cdot \vec B = 0, \\
&
\label{1d}
\nabla^2 \vec B =  P_m(\partial_t \vec B + \vec u \cdot \nabla \vec B
-  \vec B \cdot \nabla \vec u),
\end{align}
\end{subequations}
where $\partial_t$ denotes the partial derivative with respect to time
$t$ and where all terms in the equation of motion that can be written
as gradients have been combined into $ \nabla \pi$. The Boussinesq
approximation has been assumed in that the density $\varrho$ is
regarded as constant except in the gravity term where its temperature
dependence given by $\alpha \equiv - ( \de \varrho/\de T)/\varrho =${\sl
const} is taken into account. The Rayleigh number $R$,
the Coriolis number $\tau$, the Prandtl number $P$ and the magnetic
Prandtl number $P_m$ are defined by 
\begin{equation}
R = \frac{\alpha \gamma \beta d^6}{\nu \kappa} , 
\enspace \tau = \frac{2
\Omega d^2}{\nu} , \enspace P = \frac{\nu}{\kappa} , \enspace P_m = \frac{\nu}{\lambda},
\end{equation}
where $\lambda$ is the magnetic diffusivity.  Because the velocity 
field $\vec u$ as well as the magnetic flux density $\vec B$ are
solenoidal vector fields,   the general representation in terms of
poloidal and toroidal components can be used 
\begin{subequations}
\begin{align}
&
\vec u = \nabla \times ( \nabla v \times \vec r) + \nabla w \times 
\vec r \enspace , \\
&
\vec B = \nabla \times  ( \nabla h \times \vec r) + \nabla g \times 
\vec r \enspace .
\end{align}
\end{subequations}
By multiplying the (curl)$^2$ and the curl of equation \eqref{1a} by
$\vec r$ we obtain two equations for $v$ and $w$  
\begin{subequations}
\label{momentum}
\begin{align}
&\hspace*{-1.6cm}
[( \nabla^2 - \partial_t) {\cal L}_2 + \tau \partial_{\varphi} ] \nabla^2 v +
\tau {\cal Q} w - {\cal L}_2 \Theta  
= - \vec r \cdot \nabla \times [ \nabla \times ( \vec u \cdot
\nabla \vec u - \vec B \cdot \nabla \vec B)], \\
&\hspace*{-1.6cm}
[( \nabla^2 - \partial_t) {\cal L}_2 + \tau \partial_{\varphi} ] w - \tau {\cal Q}v 
= \vec
r \cdot \nabla \times ( \vec u \cdot \nabla \vec u - \vec B \cdot
\nabla \vec B), 
\end{align}
\end{subequations}
where $\partial_{\varphi}$ denotes the partial derivative with respect to
the angle $\varphi$ of a spherical system of coordinates $r, \theta, \varphi$
and where the operators ${\cal L}_2$ and $\cal Q$ are defined by 
\begin{align}
&
{\cal L}_2 \equiv - r^2 \nabla^2 + \partial_r ( r^2 \partial_r), \nonumber\\
&
{\cal Q} \equiv r \cos \theta \nabla^2 - ({\cal L}_2 + r \partial_r ) ( \cos \theta
\partial_r - r^{-1} \sin \theta \partial_{\theta}). \nonumber
\end{align}
The heat equation for the dimensionless deviation $\Theta$ from the
static temperature distribution can be written in the form
\begin{equation}
\label{heat}
\nabla^2 \Theta + R{\cal L}_2 v = P ( \partial_t + \vec u \cdot \nabla ) \Theta,
\end{equation}
and the equations for $h$ and $g$ are obtained through the multiplication of
equation \eqref{1d} and of its curl by $\vec r$
\begin{subequations}
\label{induction}
\begin{align}
&
\nabla^2 {\cal L}_2 h = P_m [ \partial_t {\cal L}_2 h - \vec r \cdot
\nabla \times ( \vec u \times \vec B )], \\
&
\nabla^2 {\cal L}_2 g = P_m [ \partial_t {\cal L}_2 g - \vec r \cdot
\nabla \times ( \nabla \times ( \vec u \times \vec B ))].
\end{align}
\end{subequations}
We assume stress-free boundaries with fixed temperatures and use the value 0.4 for the 
radius ratio $\eta=r_i/r_o$,
\begin{equation}
\label{vbc}
\hspace*{-8mm}
v = \partial^2_{rr}v = \partial_r (w/r) = \Theta = 0 
\qquad \mbox{ at } r=r_i \equiv 2/3  \mbox{ and } r=r_o \equiv 5/3.
\end{equation}
For the magnetic field electrically insulating
boundaries are assumed such that the poloidal function $h$ must be 
matched to the function $h^{(e)}$ which describes the  
potential fields 
outside the fluid shell  
\begin{equation}
\hspace*{-8mm}
\label{mbc}
g = h-h^{(e)} = \partial_r ( h-h^{(e)})=0 \qquad 
\qquad \mbox{ at } r=r_i \equiv 2/3  \mbox{ and } r=r_o \equiv 5/3.
\end{equation}
But computations for the case of an inner boundary with no-slip
conditions and an electrical conductivity equal to that of the fluid
have also been done. The numerical integration of equations
\eqref{momentum},\eqref{heat} and \eqref{induction} together with boundary
conditions \eqref{vbc} and \eqref{mbc} proceeds with the pseudo-spectral 
method as described by Tilgner and Busse (1997)
which is based on an expansion of all dependent variables in
spherical harmonics for the $\theta , \varphi$-dependences, i.e. 
\begin{equation}
v = \sum \limits_{l,m} V_l^m (r,t) P_l^m ( \cos \theta ) \exp \{ im \varphi \}
\end{equation}
and analogous expressions for the other variables, $w, \Theta, h$ and $g$. 
$P_l^m$ denotes the associated Legendre functions.
For the $r$-dependence expansions in Chebychev polynomials are used. 
For further details see also Busse {\it et al.} (1998) or Grote {\it
  et al.} (1999).
For the computations to be reported in the following a minimum of
33 collocation points in
the radial direction and spherical harmonics up to the order 64 have been
used. But in many cases the resolution has been increased to 49 collocation
points and spherical harmonics up to the order 96 or 128.
\begin{figure}
\vspace*{4mm}
\begin{center}
\epsfig{file=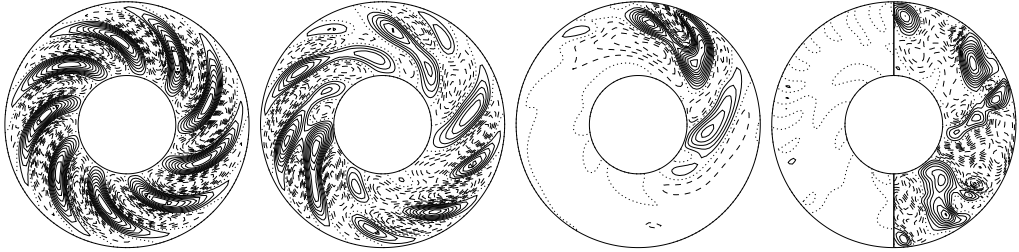,width=32pc,clip=} 
\end{center}
\caption[]{Equatorial streamlines $r \dd_\varphi v=$ const.\ in the
  case  $P=0.5$, $\tau=10^4$ and $R=1.8\x10^5$
 $3.2 \x 10^5$, $4 \x 10^5$, $9\x 10^5$ (from left to right). 
 The two halves of last plot show convection in the minimum and in the
 maximum of a relaxation cycle.} 
\label{f.02}
\end{figure}

\section{Convection in rotating spherical shells}

\begin{figure}
\vspace*{4mm}
\begin{center}
\epsfig{file=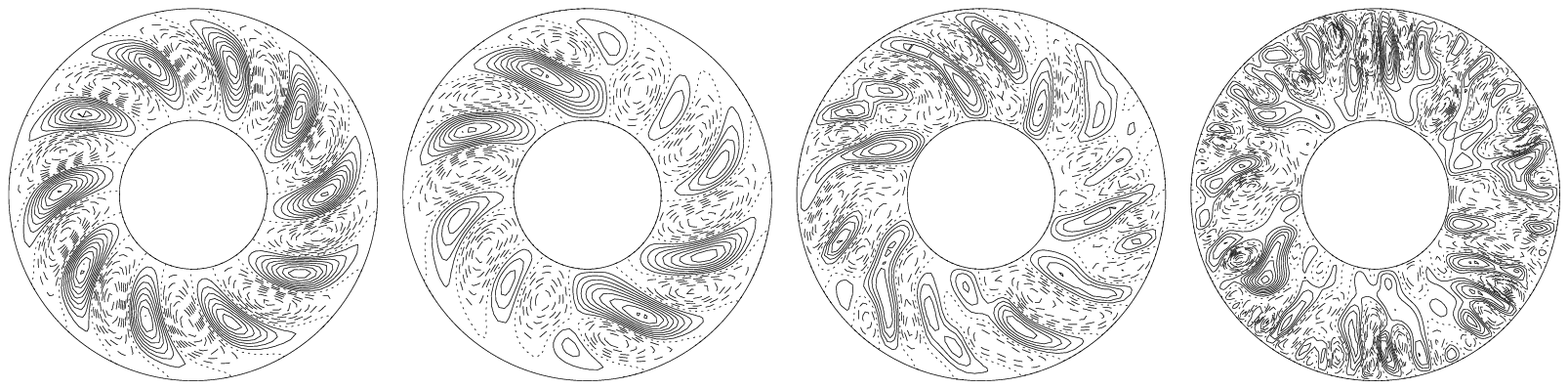,width=32pc,angle=0,clip=} 
\end{center}
\caption{Equatorial streamlines $r \dd_\varphi v=$ const.\ in
the case  $P=20$, $\tau=5\x10^3$ and  $R=1.6
\x 10^5$, $1.75 \x 10^5$, $2.5 \x 10^5$, $1.5\x10^6$ (from left to right).}
\label{f.03}
\vspace*{4mm}
\begin{center}
\epsfig{file=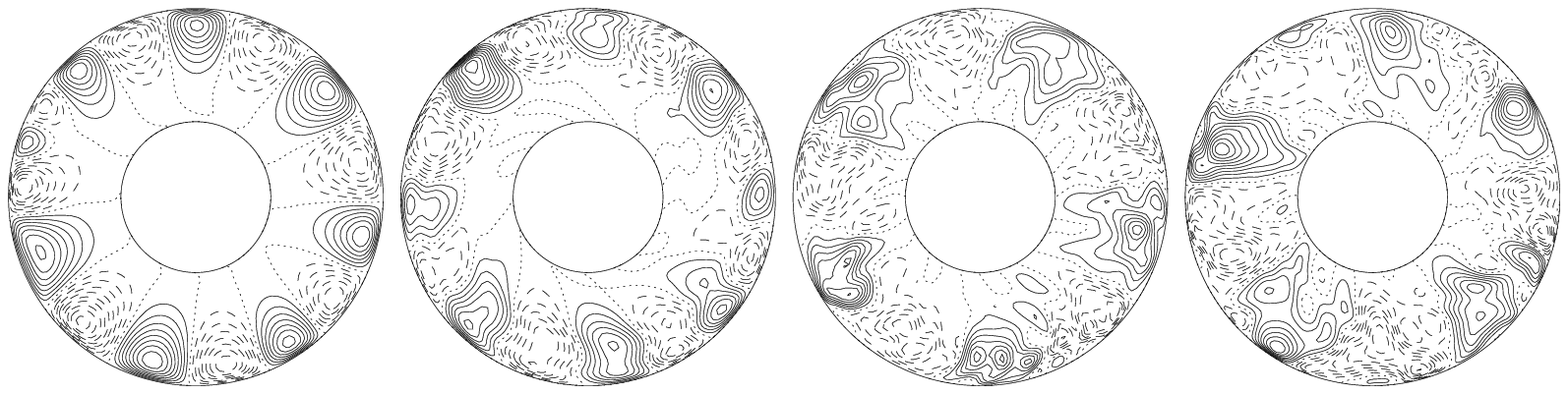,width=32pc,angle=0,clip=} 
\end{center}
\caption[]{Equatorial streamlines $r \dd_\varphi v=$ const.\ in the
  case $P=0.025$, $\tau=10^5$ and $R=4\x10^5$,
 $6 \x 10^5$, $8 \x 10^5$, $10^6$ (from left to right).} 
\label{f.04}
\end{figure}
Three different types of convection can be distinguished in rotating
spherical shells. Predominantly convection occurs in the form of rolls
aligned with the axis of rotation which exhibit properties of thermal
Rossby waves in that they are drifting in the prograde azimuthal
direction. They are confined to the region outside the virtual surface
of the tangent cylinder which touches the inner boundary at its
equator. The dynamics of these convection columns, as they are
sometimes called, is intimately connected for Prandtl numbers of the
order unity or less with the differential rotation that is generated
by their Reynolds stresses. Above their onset as a $m$-periodic
pattern in the azimuthal direction the convection columns experience
with increasing Rayleigh number transitions to amplitude and shape
vacillations before they become spatio-temporally chaotic in the
dimensions perpendicular to the axis while retaining their nearly
perfect alignment with the rotation vector. In this regime of
beginning turbulence coherent processes such as localized convection
and relaxation oscillations are realized through the interaction of
the differential rotation and the convection columns. A graphical
display of these stages of convection is shown in figure \ref{f.02} in terms of
the streamlines of the convection columns intersected by the
equatorial plane. For a more detailed description we refer to Grote
and Busse (2001) or the review of Busse (2002a). As the Reynolds
stresses decrease with increasing $P$ these coherent processes
disappear. Figure \ref{f.03} indicates that the spiralling nature of the
convection columns also diminishes with increasing Rayleigh number. A
small non-geostrophic differential rotation persist driven as a
"thermal wind" by temperature gradients caused by the lateral
inhomogeneity of the convective heat transport.  
\begin{figure}
\vspace*{4mm}
\begin{center}
\epsfig{file=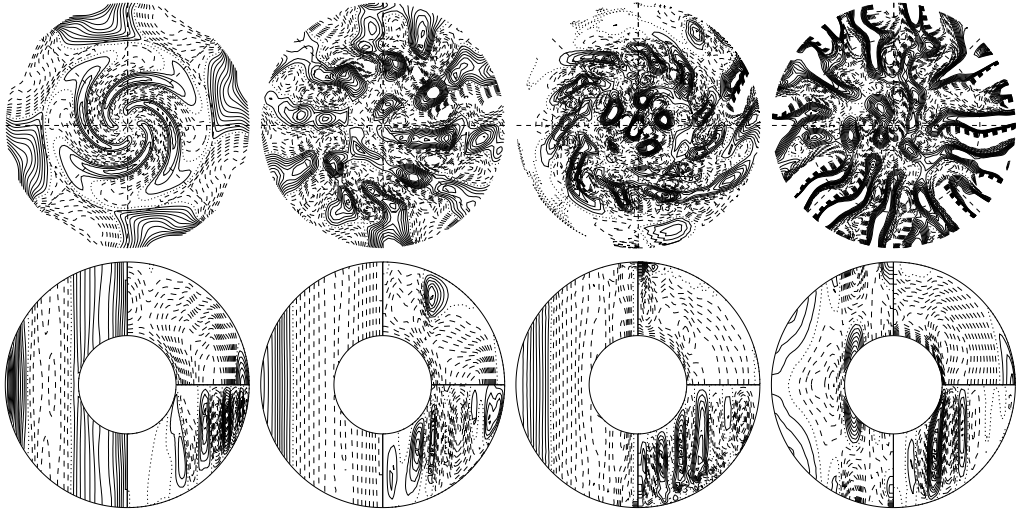,width=32pc,angle=0,clip=} 
\end{center}
\caption{Polar convection in the cases, $P=0.025$,
  $\tau=5\x10^4$, $R=2\x10^5$;  
  $P=0.1$, $\tau=3\x10^4$, $R=7.5\x10^5$; $P=1$, $\tau=10^4$,
  $R=1.4\x10^6$ and $P=20$, $\tau=5\x10^3$, $R=10^6$ (from left to
  right). The plots in the upper row show lines of constant
  $u_r$ at the surface $r=r_i+0.5$ as seen from the
  north pole. Isolines corresponding
  to higher amplitude have not been plotted in order to emphasize
  structures with lower amplitude.
  The plots  in the lower row exhibit lines of constant
  $\overline{u}_\varphi$ in the left half, of $\overline{\Theta}$ in
  the upper right quarter and of the streamlines $r \sin \theta
  \dd_\theta \overline{v} $ in
  the lower right quarter, all in the meridional plane.}
\label{f.05}
\end{figure}

A second distinct form of convection are the equatorially attached
cells which represent modified inertial modes and become the preferred
form of convection at sufficiently low values of $P$. They were first
found by Zhang and Busse (1987) and analytical descriptions in terms
of perturbed inertial oscillations have been given by Zhang (1994,
1995) and Busse and Simitev (2004a). The equatorially attached cells do
not develop strong Reynolds stresses and they are thus less subject to
the disruptive effects of the shear of a differential rotation. But
the continuity of the convective heat transport requires that the
equatorially attached convection occurs in conjunction with the
columnar convection closer to the inner boundary of the shell. This
effect is visible in the form of secondary extrema of the streamlines
far from the boundary in figure \ref{f.04} where the evolution of the
convection flows in the chaotic regime is illustrated. Note that the
strong attachment to the outer equatorial boundary persists even at
the highest value of $R$ used in this figure. For further details see 
Simitev and Busse (2003).
\begin{figure}
\vspace*{4mm}
\mypsfrag{1}{1}
\mypsfrag{aaa1}{}
\mypsfrag{aaa2}{\hspace*{1mm}$10^0$}
\mypsfrag{aaa3}{}
\mypsfrag{aaa4}{\hspace*{1mm}$10^2$}
\mypsfrag{aaa5}{}
\mypsfrag{aaa6}{\hspace*{1mm}$10^4$}
\mypsfrag{baa1}{$10^{-2}$}
\mypsfrag{baa2}{}
\mypsfrag{baa3}{\hspace*{1mm}$10^0$}
\mypsfrag{baa4}{}
\mypsfrag{baa5}{\hspace*{1mm}$10^2$}
\mypsfrag{baa6}{}
\mypsfrag{baa7}{\hspace*{1mm}$10^4$}
\mypsfrag{caa1}{}
\mypsfrag{caa2}{\hspace*{1mm}$10^1$}
\mypsfrag{caa3}{}
\mypsfrag{caa4}{\hspace*{1mm}$10^3$}
\mypsfrag{caa5}{}
\mypsfrag{caa6}{\hspace*{1mm}$10^5$}
\mypsfrag{daa1}{}
\mypsfrag{daa2}{$10^{-3}$}
\mypsfrag{daa3}{}
\mypsfrag{daa4}{}
\mypsfrag{daa5}{}
\mypsfrag{daa6}{\hspace*{1mm}$10^{1}$}
\mypsfrag{eaa1}{\hspace*{1mm}$10^1$}
\mypsfrag{eaa2}{\hspace*{1mm}$10^2$}
\mypsfrag{eaa3}{\hspace*{1mm}$10^3$}
\mypsfrag{eaa4}{\hspace*{1mm}$10^4$}
\mypsfrag{faa1}{\hspace*{1mm}$10^2$}
\mypsfrag{faa2}{}
\mypsfrag{faa3}{\hspace*{1mm}$10^4$}
\mypsfrag{1.0}{1}
\mypsfrag{3.0}{3} 
\mypsfrag{5.0}{5} 
\mypsfrag{2.0}{} 
\mypsfrag{4.0}{} 
\mypsfrag{2.3}{2.3} 
\mypsfrag{1.00}{1}
\mypsfrag{1.02}{1.02}
\mypsfrag{1}{1}
\mypsfrag{10}{\hspace*{-2mm}10}
\mypsfrag{0}{\hspace*{-2mm}0.1}
\mypsfrag{8.0}{8} 
\mypsfrag{15.0}{15} 
\mypsfrag{1.0}{1} 
\mypsfrag{1.2}{1.2} 
\mypsfrag{1.4}{1.4} 
\mypsfrag{1.00}{1}
\mypsfrag{1.07}{1.07}
\mypsfrag{1.14}{1.14}
\mypsfrag{a}{\sl(a)}
\mypsfrag{b}{\sl(b)}
\mypsfrag{c}{\hspace*{-1mm}\sl(c)}
\mypsfrag{d}{\sl(d)}
\mypsfrag{R}{\hspace*{-8mm}$(R-R_c)/R_c$}               
\begin{center}
\begin{tabular}{@{}c@{\extracolsep{-4mm}}c@{}}
\epsfig{file=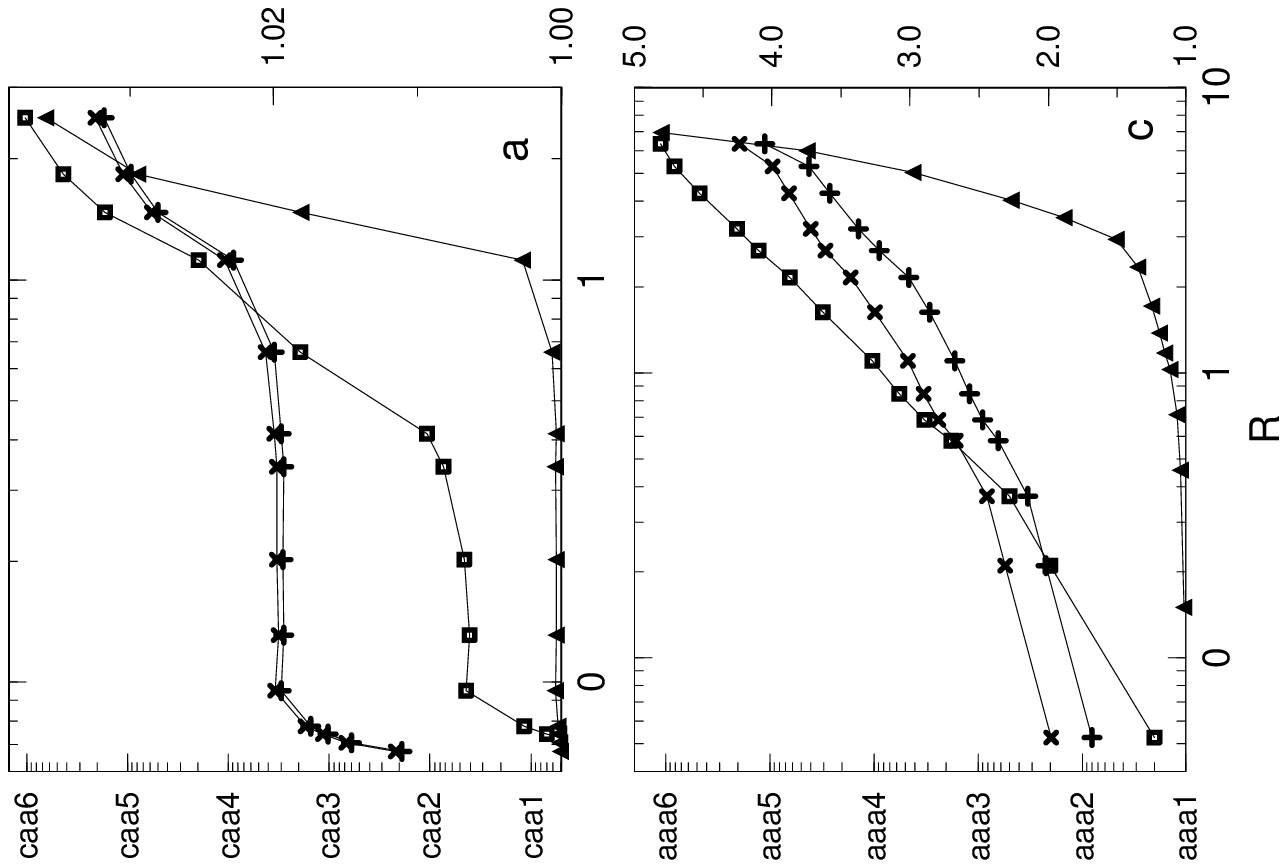,height=7cm,angle=-90} &
\epsfig{file=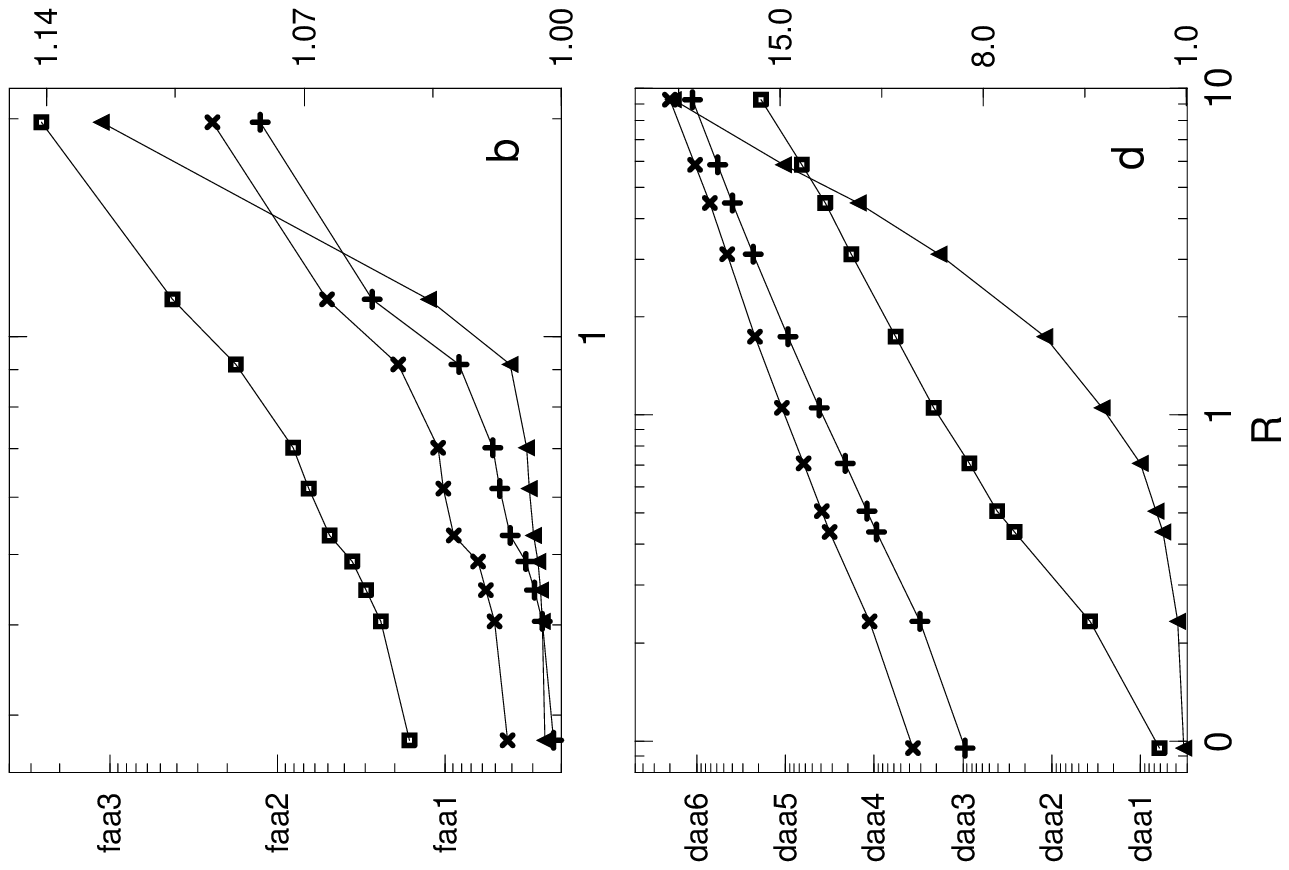,height=7cm,angle=-90} \\
\end{tabular}
\end{center}
\hspace*{-3mm}
\caption[]{Time-averaged energy densities $\overline{E}_t$ (squares),
  $\check E_p$ (plus-signs), $\check E_t$ (crosses)  and
  Nusselt number $Nu_i$ (filled triangles,  right ordinates) as
  functions of $R/R_c$ for  
{\sl(a)} $P =0.025$, $\tau = 10^5$,
{\sl(b)} $P =0.1$, $\tau = 3\x10^4$,
{\sl(c)} $P=1$, $\tau = 10^4$, and
{\sl(d)} $P =20$, $\tau = 5 \x10^3$. The values of $R_c$
are $283000$, $235000$, $190000$ and $146000$ in the cases {\sl (a)},
{\sl (b)}, {\sl (c)} and {\sl (d)} respectively.
The mean poloidal energy component $\overline{E}_p$ has been
omitted since it is by factor $10^3$--$10^2$ smaller than the other
components.}
\label{f.06}
\vspace{-5mm}
\end{figure}

A third form of convection is realized in the polar regions of the
shell which comprise the two fluid domains inside the tangent
cylinder. Since gravity and rotation vectors are nearly parallel in
these regions (unless large values of $\eta$ are used) convection
resembles the kind realized in a horizontal layer heated from below
and rotating about a
vertical axis. A tendency towards an alignment of convection rolls
with the North-South direction (Busse and Cuong, 1977) can be noticed,
but this property is superseded by instabilities of the
K\"uppers-Lortz type and by interactions with turbulent convection
outside the tangent cylinder. 
The onset of convection in the polar regions generally occurs at
Rayleigh numbers considerably above the critical values $R_c$ for
onset of convection outside the tangent cylinder. 
Except for the case of very low Prandtl numbers the differential rotation in the polar regions is usually
oriented in the direction opposite to that of rotation and thus tends
to facilitate polar convection by reducing the rotational constraint.
The possibility
exists, however, that at sufficiently low values of $P$ and high
values of $\tau$ finite amplitude convection in the polar regions may
precede the onset of convection in other regions. 
\begin{figure} \vspace*{4mm}
\mypsfrag{a2}{\hspace*{-2mm}$10^2$}
\mypsfrag{Rm}{\hspace*{-2mm}$Rm$}
\npsfrag{0}{tl}{0.1}  
\npsfrag{1}{tl}{1}  
\npsfrag{10}{tl}{10}  
\npsfrag{Pm}{tl}{$P_m$}
\begin{center}
\hspace*{-4mm}
\epsfig{file=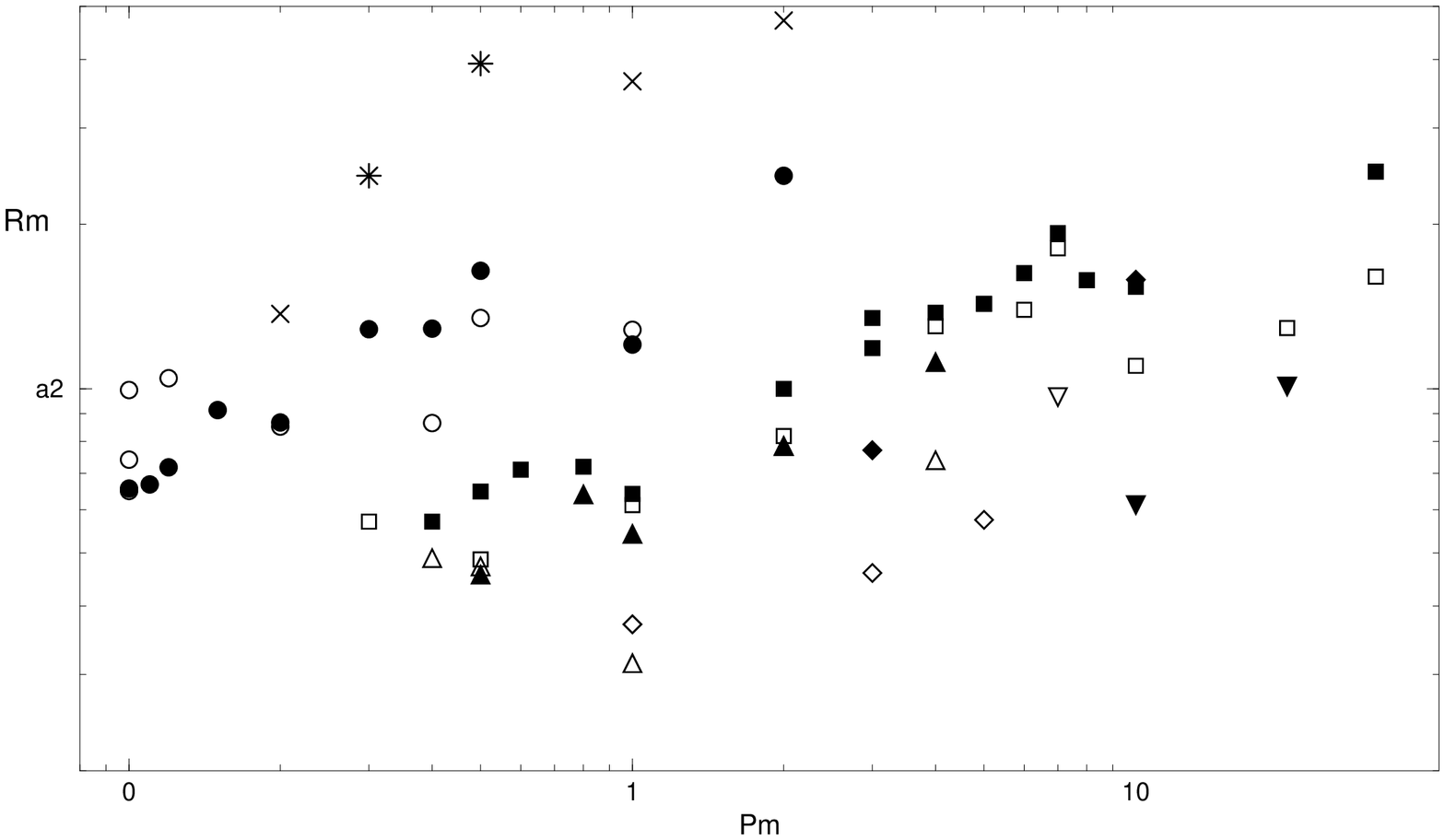,width=6cm,angle=-90}
\end{center}
\vspace*{2mm}
\caption[]{Magnetic Reynolds numbers $Rm$ for the
  onset of dynamo action as a function of $P_m$ in the cases 
  $P=0.01$, $\tau=10^5$ (stars), $P=0.025$, $\tau=10^5$ (crosses), 
$P=0.1$,  $\tau=10^5$ (circles),  $P=1$, $\tau=3\x10^4$ (triangles up), $P=1$,
  $\tau=10^4$ (squares)  and  $P=5$,
  $\tau=5\x10^3$ (diamonds) and $P=10$, $\tau=5\x10^3$ (triangles
  down). The empty symbols in the cases with $P \ge 0.1$ are based on decaying 
  dynamos, the full symbols are based on the lowest non-decaying solutions.} 
\label{f.07}
\end{figure}
The patterns of polar convection and other properties discussed in this
section are illuminated by the display of convection at different
values of $P$ in figure \ref{f.05}. 

Quantitative aspects of convection are
described by figure \ref{f.06} where the averages over space and time of the
kinetic energy densities of the various components of the convection
flow are shown as a function of $R$. The energy densities are defined by
\begin{subequations}
\label{edens}
\begin{align}
&
\overline{E}_p = \frac{1}{2} \langle \mid \nabla \times ( \nabla \bar v \times \vec r )
\mid^2 \rangle , \quad \overline{E}_t = \frac{1}{2} \langle \mid \nabla \bar w \times
\vec r \mid^2 \rangle, \\
&
\check{E}_p = \frac{1}{2} \langle \mid \nabla \times ( \nabla \check v \times \vec r )
\mid^2 \rangle , \quad \check{E}_t = \frac{1}{2} \langle \mid \nabla \check w \times
\vec r \mid^2 \rangle,
\end{align}
\end{subequations}
where the angular brackets indicate the average over the fluid shell
and $\bar v$ refers to the azimuthally averaged component of $v$,
while $\check v$ is defined by $\check v = v - \bar v $. The Nusselt
number at the inner spherical boundary $Nu_i$ is also shown in figure
\ref{f.06}. It is defined by
\begin{equation}
Nu_i=1- \frac{P}{r_i} \left.\frac{\de \overline{\overline{\Theta}}}{\de r}\right|_{r=r_i}
\end{equation}  
where the double bar indicates the average over the spherical surface.
The rapid growth with $R$ of $\overline{E}_t$ corresponding to the energy of
differential rotation is remarkable in the cases {\sl (a)}, {\sl (b)},
and {\sl (c)}  of figure \ref{f.06}. Only for higher values of $P$ as
in case {\sl (d)} does $\overline{E}_t$ 
never exceed the energies of the fluctuating components of
motion. Another remarkable feature of the plots is the rapid rise of
the Nusselt number $Nu_i$ after the onset of amplitude vacillations and
the transition to a chaotic time dependence. The steadily drifting
convection columns or the equatorially-attached cells do not transport
heat very well because of the mismatch between the geometry of the
boundaries and the flow structure. Because this mismatch does not
occur in the case of polar convection, the heat transport of the
latter may easily exceed that of convection outside the tangent
cylinder at higher values of $R$. 
The dependence of the kinetic energy densities on $R$ in case {\sl
(a)} of figure \ref{f.06} are somewhat unusual in that after a rapid
growth near onset a regime of saturation follows and a growth of the
kinetic energies with $R$ begins again only when $R$ has reached
nearly twice
its critical value. The behavior is caused by the competition of
several modes with different azimuthal wavenumbers $m$ as is described
in more detail by Simitev and Busse (2003). The ways in which
properties of convection are reflected in the generated magnetic
fields will be discussed in the following sections. 

\section{The onset of dynamo action for different Prandtl numbers}

\begin{figure} \vspace*{4mm}
\mypsfrag{o}{\hspace*{-.8mm}$\odot$}
\psfrag{D}{\hspace*{-.8mm}$\blacktriangle$}
\psfrag{Q}{\hspace*{-1.2mm}$\bigstar$}
\mypsfrag{H}{\hspace*{-.8mm}$\triangle$}
\mypsfrag{M}{\hspace*{-.4mm}$\blacksquare$}
\mypsfrag{P}{$P$}
\mypsfrag{Pm}{$P_m$}
\mypsfrag{R}{$R\x10^{-5}$}
\mypsfrag{b0}{0} 
\mypsfrag{b5}{5} 
\mypsfrag{b10}{10}
\mypsfrag{b15}{15}
\mypsfrag{b20}{20}
\mypsfrag{1}{1}  
\mypsfrag{3}{3}  
\mypsfrag{5}{5}  
\mypsfrag{10}{10} 
\mypsfrag{20}{20}
\mypsfrag{50}{50}
\npsfrag{a1}{tl}{1}  
\npsfrag{a3}{tl}{3}  
\npsfrag{a5}{tl}{5}  
\npsfrag{a10}{tl}{10}
\npsfrag{a20}{tl}{20}
\npsfrag{a50}{tl}{50} 
\npsfrag{a100}{tl}{100}
\npsfrag{a200}{tl}{200}
\begin{center}
\hspace*{15mm}
\epsfig{file=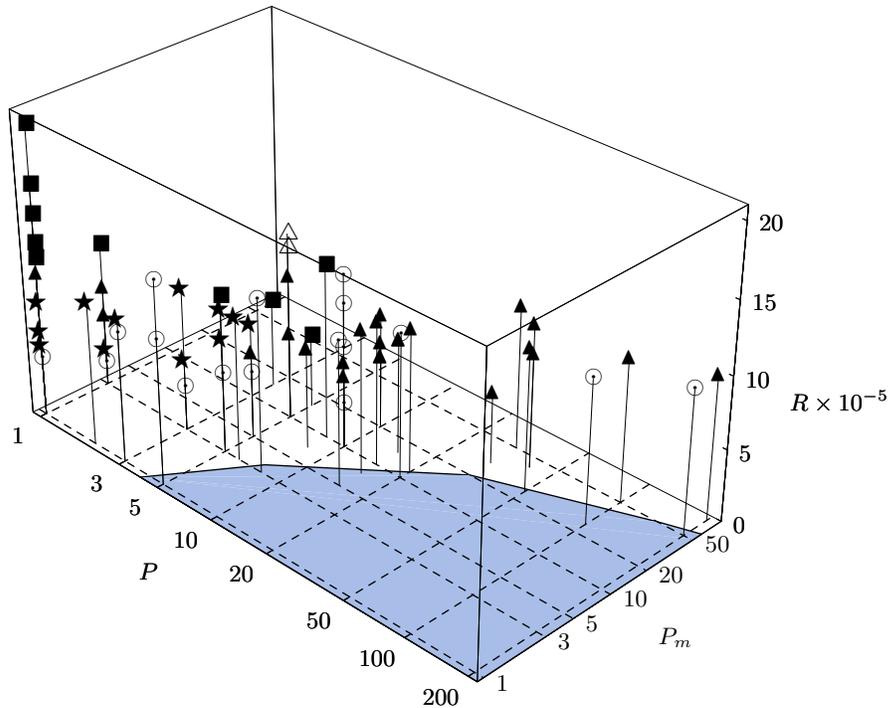,width=11cm}
\end{center}
\vspace*{-3mm}
\caption[]{Convection driven dynamos as a function of $R$, 
  $P$ and $P_m$ for  $\tau=5\x10^3$. The symbols indicate chaotic 
  dipolar ($\blacktriangle$), hemispherical ($\triangle$),
  quadrupolar ($\bigstar$), mixed ($\blacksquare$) and decaying dynamos ($\odot$).}
\label{f.08}
\end{figure}
The onset of convection driven dynamos in the case $P = 1$ has been
explored in numerous papers in the past. We refer to the work of
Kageyama and Sato (1997), Kida and Kitauchi (1998), Ishihara and Kida
(2002), Busse \etal~(1998), Christensen \etal~(1999), Olson \etal~
(1999), Katayama \etal~(1999), Takahashi \etal~(2003), Grote
\etal~(2000), Grote and 
Busse (2001). In particular the results of the latter two papers can
be compared well with the results presented in the following since
only the Prandtl number differs among the parameters of the
computations. Since one of the goals of the numerical simulations is
the search for dynamos with low values of the magnetic Prandtl number
$P_m$, the amplitude of convection must be sufficiently high such that
the magnetic Reynolds number defined by $Rm = (2E)^{1/2}P_m$ reaches a
value of the order $10^2$. Here $E$ refers to the average density of
the kinetic energy of convection, i.e. $E = \overline{E}_p +
\overline{E}_t + \check{E}_p + \check{E}_t$. In figure \ref{f.07}
values of $Rm$ for cases of sustained dynamo 
action and for cases of decaying magnetic fields have been indicated
for several values of $P$. No special effort has been made to
determine the minimum value $Rm_{min}$ of $Rm$ for dynamo
action. Since in all cases convection is chaotic and since dynamos are
typically subcritical, \ie the boundary between sustenance of a
dynamo and its decay occurs at finite amplitudes of the magnetic field,
$Rm_{min}$ is not a well defined quantity. Moreover, the value 
$Rm$ as defined above will decrease after the onset of dynamo action
since typically the energy of differential rotation is strongly
reduced by the action of the Lorentz force. 

It is evident from figure \ref{f.07} that the goal of low values $P_m$
is most readily reached at 
low Prandtl numbers. At high values of $P$ the minimum value of the 
magnetic Reynolds numbers for dynamos decreases slightly, but $P_m$ must 
increase in proportion to $P$ in order to achieve a finite ratio
$P/P_m$ in the limit of infinite Prandtl number as shown by Zhang
and Busse (1990).
The tendency of low Prandtl numbers to lower the minimum values of
$P_m$ required for dynamos is limited, however, by the transition to
inertial convection for $P\le(\tau/\tau_0)^{-2}$ where $\tau_0$ is a
quantity of the order $10^2$ (Ardes et al., 1997). It thus has not
been possible to obtain dynamos with values of $P_m$ below $0.1$ for
Prandtl numbers less than $0.05$. Inertial convection is evidently
less conducive to dynamo action than columnar convection and higher
values of $R_m$ are required for dynamos at $P=0.025$ and $P=0.01$
than at higher Prandtl numbers according to figure \ref{f.07}. The most
promising way towards low $P_m$ dynamos thus appears to be an
increasing $\tau$ with an accompanying reduction of $P$. Of course,
the numerical resolution will have to be increased at the same time.  
\begin{figure} \vspace*{4mm}
\npsfrag{0.1}{tl}{0.1}
\npsfrag{1.0}{tl}{1.0}
\npsfrag{Pm}{tl}{$P_m$}
\mypsfrag{R}{$R$}
\mypsfrag{aaa6}{\hspace{-1mm}$10^{6}$}
\mypsfrag{aaa7}{\hspace{-1mm}$10^7$}
%
\begin{center}
\hspace{-1cm}
\epsfig{file=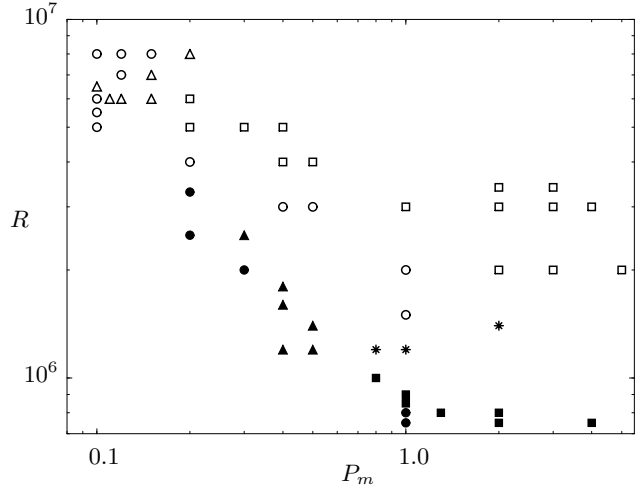,width=6cm,angle=-90}
\vspace*{2mm}
\end{center}
\caption[]{Convection driven dynamos as a function of the  Rayleigh
  number $R$ and the magnetic Prandtl number  $P_m$ for $P=0.1$,
  $\tau=10^5$ (empty symbols) and $\tau=3\x10^4$ (filled symbols). 
  The symbols indicate chaotic dipolar (squares), hemispherical
  (triangles), mixed (stars) and decaying dynamos  (circles).}  
\label{f.09}
\end{figure}
\begin{figure} \vspace*{4mm}
\mypsfrag{Md}{\hspace*{-4mm}$M_{dip}$}
\mypsfrag{Mq}{\hspace*{-4mm}$M_{quad}$}
\mypsfrag{Es}{\hspace*{-1mm}$E$}
\mypsfrag{0}{0}
\mypsfrag{5}{5}
\mypsfrag{10}{10}
\mypsfrag{1}{1}
\mypsfrag{2}{2}
\npsfrag{t}{tl}{$t$}
\mypsfrag{x}{\hspace*{-4mm}$\x10^4$}
\npsfrag{0.05}{tl}{0.05}
\npsfrag{0.07}{tl}{0.07}
\npsfrag{0.09}{tl}{0.09}
\npsfrag{0.11}{tl}{0.11}
\npsfrag{0.13}{tl}{0.13}
\begin{center}
\hspace*{-8mm}
\epsfig{file=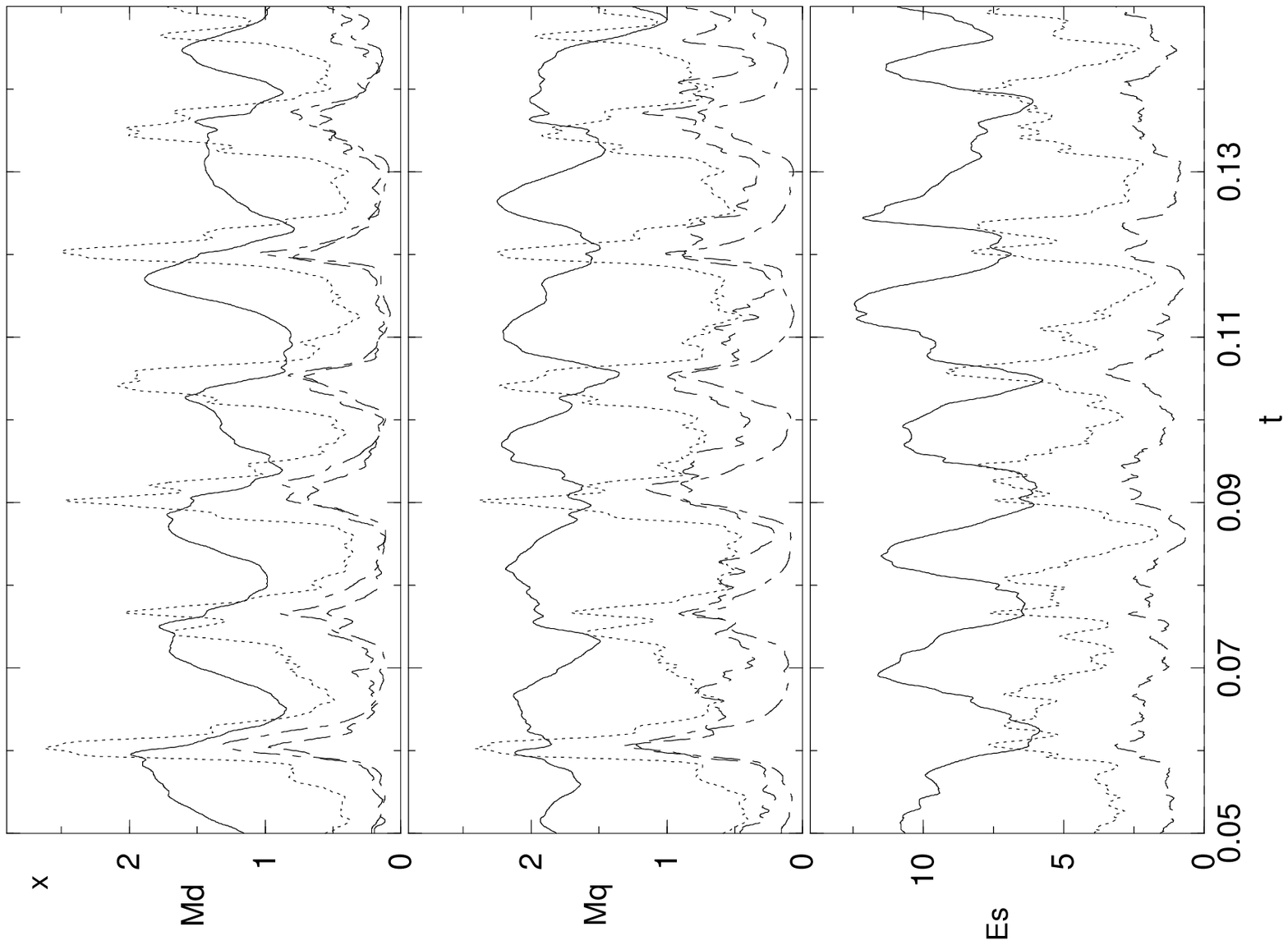,height=8cm,width=10cm,angle=-90}
\end{center}
\vspace*{2mm}
\caption[]{Time series of energy densities of a hemispherical dynamo
  in the case $P=0.1$, $\tau=10^5$, $R=6\x10^6$, $P_m=0.11$. 
  The upper and middle panels show energy densities of
  dipolar and quadrupolar components of the magnetic field, while the
  lower panel shows energy densities of the velocity field.
  The mean toroidal components are represented by solid lines, the
  fluctuating toroidal - by dotted lines, the mean poloidal - by
  dot-dashed lines and the fluctuating poloidal by dashed lines.}
\label{f.10}
\end{figure}

A particular feature that is evident from figure \ref{f.07} 
(and also figure \ref{f.09}) is that high
values of $R$ may prevent dynamo action. In the case $P=0.1$,
$\tau=10^5$ it is found that only a finite interval of convection
energy permits dynamo action since high values of $Rm$ lead to the
expulsion of magnetic flux from the convection eddies which is
detrimental for the generation of magnetic fields. The same effect can
also be noticed for higher values of the Prandtl number as
demonstrated in the case of $P_m=P=10$ of figure \ref{f.08}. 

A more complete overview of dynamos at values of $P$ and $P_m$ above
unity is shown in figure \ref{f.08} where the structure of the
magnetic field is also indicated. 
As long as the properties of convection are nearly symmetric with
respect to the equatorial plane three types of dynamos can be
distinguished: Quadrupolar dynamos are characterized by a magnetic
field that exhibits the same symmetry with respect to the equatorial
plane as the convection velocity field, while the opposite symmetry
characterizes dipolar dynamos. Hemispherical dynamos correspond to the
superposition of a quadrupolar and a dipolar magnetic field of nearly
the same form, but with opposite signs in the northern and the
southern hemispheres such that the magnetic field nearly vanishes in
one of the hemispheres (Grote and Busse, 2000). Once the symmetry of
convection with respect to the equatorial plane is impaired
significantly, -- as happens, for example, at higher Rayleigh numbers
through the onset of polar convection --, the structures of dynamos can
no longer be clearly distinguished and we speak of ``mixed''
dynamos. These dynamos continue to exhibit an alignment with the axis
of rotation. Dynamos with a dominant equatorial dipole, for instance,
have not been found.

For a fixed value of $P$ quadrupolar
dynamos are obtained for sufficiently low values of $P_m$ which give way
to hemispherical and dipolar dynamos with increasing  $P_m$ as has also
been observed in the case $P=1$ (Grote \etal, 2000). But since dynamos
for low values of $P_m$ disappear with increasing $P$ quadrupolar and
hemispherical dynamos can no longer be obtained as $P$ exceeds a value
of the order 5 in the case of $\tau=5\x10^3$.
Quadrupolar dynamos thus appear to be restricted to Prandtl numbers of the
order unity and moderate values of $\tau$. When $\tau$ is increased
from $10^4$ to $3\x 10^4$ quadrupolar dynamos do no longer seem to
be accessible according to the results of Grote \etal\ (2001, see also
Busse, 2002a). The
same effect occurs as $P$ is lowered as indicated in figure
\ref{f.09} for the case $P = 0.1$, $\tau = 10^5$. Here the lowest value of $P_m$
corresponds to a hemispherical dynamo. 

It should be noted that the
critical value $R_c$ of the Rayleigh number increases nearly in proportion to
$\tau P$ for $P < 1$ and nominally higher values of $\tau$ can thus be
reached for lower values of $P$. The degree of chaos increases,
however, with decreasing $P$ because of the short thermal time
scale. Nevertheless coherent structures are also found such as the
relaxation oscillations apparent in the time record shown in figure
\ref{f.10}. Here the energy densities on the various components of the
magnetic field are shown which are defined in analogy to the kinetic
energy densities \eqref{edens}, 
\begin{subequations}
\begin{align}
&
\overline{M}_p = \frac{1}{2} \langle \mid\nabla \times ( \nabla \bar h \times \vec r )
\mid^2 \rangle , \quad \overline{M}_t = \frac{1}{2} \langle \mid\nabla \bar g \times
\vec r \mid^2 \rangle, \\
&
\check{M}_p = \frac{1}{2} \langle \mid\nabla \times ( \nabla \check h \times \vec r )
\mid^2 \rangle , \quad \check{M}_t = \frac{1}{2} \langle \mid\nabla \check g \times
\vec r \mid^2 \rangle.
\end{align}
\end{subequations}
In contrast to the relaxation oscillations of non-magnetic convection
mentioned in connection with figure \ref{f.02} the oscillations of
figure \ref{f.10}
have a much shorter period and originate from the oscillations of the
hemispherical magnetic field as is evident from the comparable
periods. Since the strength of the magnetic field varies more with the
phase of the cycle at low values of $P$ than at higher ones the action
of the Lorentz force on the differential rotation also exhibits
significant variations. This in turn affects the amplitude of
convection and its dynamo action and leads to the cyclic feedback
process evident in figure \ref{f.10}. We note that the computational results
displayed in figure \ref{f.10} were obtained with a resolution of 33
collocation points in the radial direction and with spherical
harmonics of the order 96. Results obtained with 41 collocation
points and spherical harmonics up to the order 108 are rather similar
and the time averaged energies differ insignificantly. 

\begin{figure} \vspace*{4mm}
\begin{center}
\epsfig{file=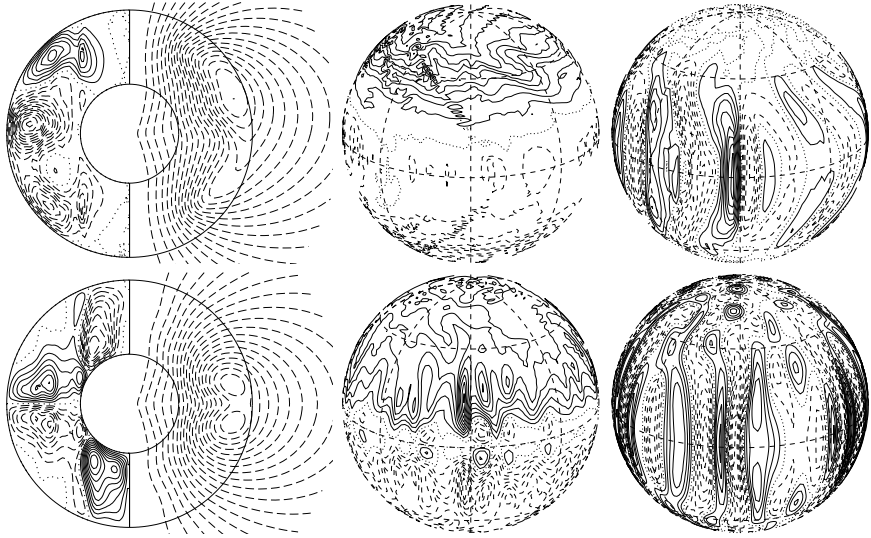,width=11.5cm,angle=0,clip=} 
\end{center}
\caption{Non-oscillatory chaotic dipolar dynamos in the cases
  $P=0.1$, $\tau=10^5$, $R=2\x10^6$, $P_m=3$ (upper row) and $P=200$, $\tau=5\x 10^3$,
  $R=10^6$, $P_m=80$ (lower row). The left column shows meridional isolines of 
  $\overline{B}_\varphi$ (left half) and of 
  $r\sin\theta \dd_\theta \overline{h}$ (right half). The middle column
  shows lines $B_r=$ const.\  at $r=r_0$. The right column shows
  lines $u_r=$ const.\  at $r=r_i+0.5$.} 
\label{f.11}
\end{figure}

Finally, we like to point out some typical differences between high and
low Prandtl number dipolar dynamos which are exhibited by the plots of
figure \ref{f.11}. At low values of $P$ the quadrupolar component 
is usually quite noticeable leading to a shift of the magnetic equator
away from the geometric one. Often this shift alternates in time in a
nearly periodic fashion. In high Prandtl number dynamos the mean
magnetic field is nearly steady and the quadrupolar component is negligible
if the Rayleigh number is not too high. This is evident in the lower
plots of figure \ref{f.11} even though the onset of polar convection with its
attendant asymmetry with respect to the equatorial plane can already
be noticed. A characteristic feature of high $P$ dynamos are the
strong zonal magnetic flux tubes in the polar regions which are mainly
caused by the thermal wind shear an example of which can be seen in
the case $P=20$ of figure 5. These flux tubes are also very evident in
the dynamo of Glatzmaier and Roberts (1995). While the differential rotation
changes a lot with increasing Prandtl number, the structure of the
convection columns hardly varies as is also evident from figure
\ref{f.11}. The rate of drift in the azimuthal direction decreases, of course,
with increasing $P$. 

\section{Energetics of dynamos}

\begin{figure} \vspace*{4mm}
\mypsfrag{0.34}{0.34}
\mypsfrag{0.39}{0.39}
\mypsfrag{0.40}{0.40}
\mypsfrag{0.00}{0.00}
\mypsfrag{0.03}{0.03}
\mypsfrag{0.07}{0.07}
\mypsfrag{0.06}{0.06}
\mypsfrag{0.08}{0.08}
\mypsfrag{16}{17}
\mypsfrag{19}{19}
\mypsfrag{22}{22}
\mypsfrag{-5}{}
\mypsfrag{0} {0} 
\mypsfrag{5} {} 
\mypsfrag{10}{10}
\mypsfrag{15}{}
\mypsfrag{20}{20}
\mypsfrag{25}{}   
\mypsfrag{-5}{}
\mypsfrag{0} {0} 
\mypsfrag{5} {} 
\mypsfrag{10}{10}
\mypsfrag{15}{}
\mypsfrag{-1.5}{-1.5}
\mypsfrag{-0.5}{-0.5}
\mypsfrag{0.5} {} 
\mypsfrag{1.5} {1.5} 
\mypsfrag{2.5} {} 
\mypsfrag{-1}  {}   
\mypsfrag{0}{0}    
\mypsfrag{1}{1}    
\mypsfrag{2}{2}    
\mypsfrag{3}{3}    
\mypsfrag{-0.5b}{} 
\mypsfrag{0.0b} {\hspace{2mm}0.0}  
\mypsfrag{0.5b} {}  
\mypsfrag{1.0b} {\hspace{2mm}1.0}  
\mypsfrag{1.5b} {}  
\mypsfrag{2.0b} {\hspace{2mm}2.0}  
\mypsfrag{-0.10}{\hspace{2mm}-0.1}
\mypsfrag{-0.05}{}
\mypsfrag{0.00}{\hspace{2mm}0.0} 
\mypsfrag{0.05}{} 
\mypsfrag{0.10}{\hspace{2mm}0.1} 
\mypsfrag{-0.4a}{} 
\mypsfrag{-0.2a}{\hspace{2.5mm}-0.2} 
\mypsfrag{0.0a}{\hspace{2mm}0.0}  
\mypsfrag{0.2a}{\hspace{2mm}0.2}   
\mypsfrag{0.0} {0}   
\mypsfrag{0.1} {}   
\mypsfrag{0.2} {0.2}   
\mypsfrag{0.3} {}   
\mypsfrag{0.4} {0.4}   
\mypsfrag{-0.2}{}  
\mypsfrag{0.0} {0.0}   
\mypsfrag{0.2} {0.2}   
\mypsfrag{0.4} {0.4}   
\mypsfrag{0.00}{0.0}  
\mypsfrag{0.02}{\hspace{2mm}.02}  
\mypsfrag{0.04}{\hspace{2mm}.04}  
\mypsfrag{0.00}{\hspace{2mm}.00}  
\mypsfrag{0.04}{\hspace{2mm}.04}  
\mypsfrag{0.08}{\hspace{2mm}.08}    
\mypsfrag{Xmp}{\hspace{7mm}$\overline{p}$}
\mypsfrag{Xmt}{\hspace{3mm}$\overline{t}$}
\mypsfrag{Xfp}{\hspace{2mm}$\check{p}$}
\mypsfrag{Xft}{\hspace{2mm}$\check{t}$}
\mypsfrag{t}{$t$}
\mypsfrag{x}{\hspace{-2mm}$\x10^7$}
\mypsfrag{x1}{\hspace{-2mm}$\x10^7$}
\hspace*{-1cm}
\epsfig{file=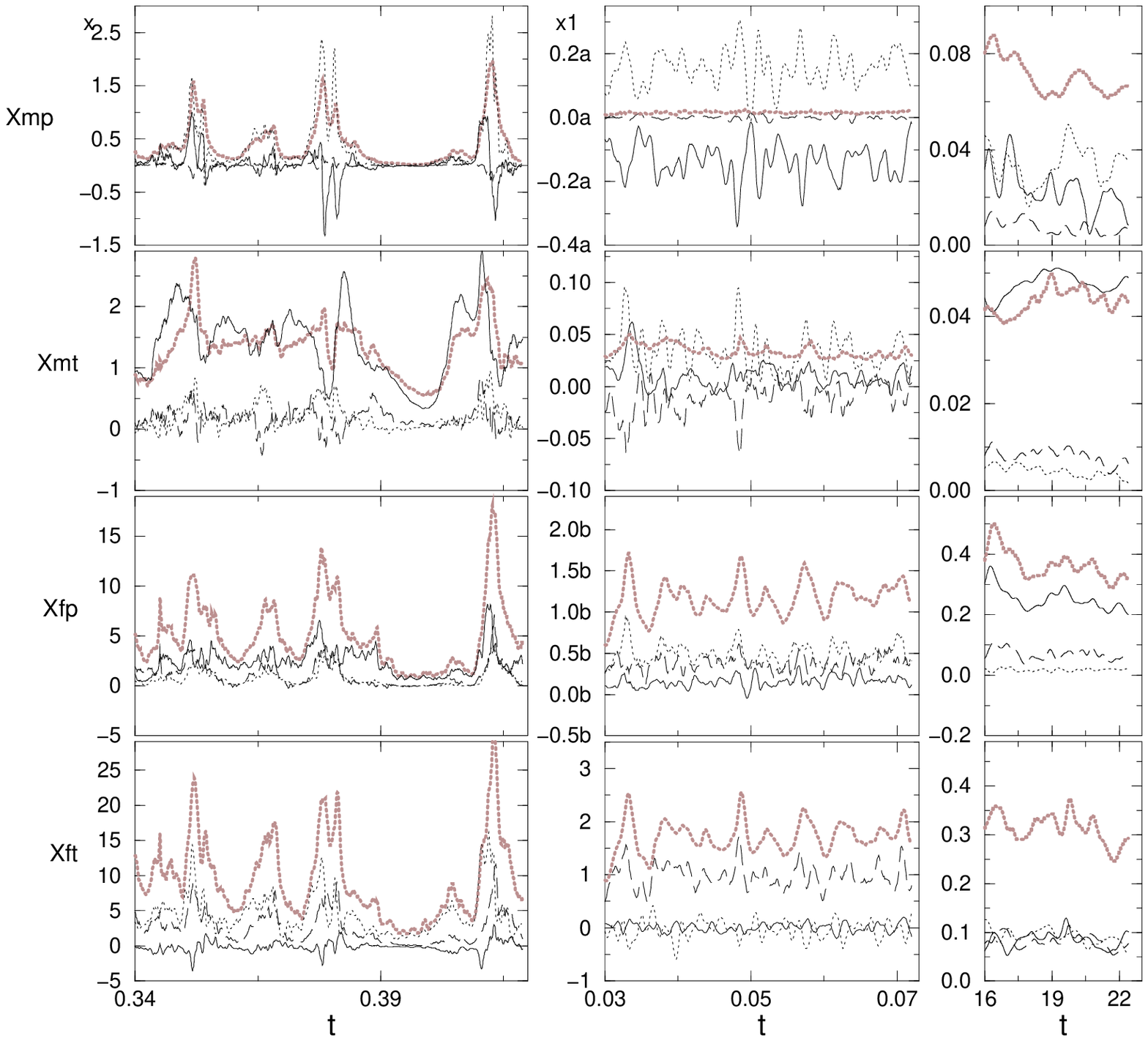,height=13cm}
\caption[]{Lorentz terms (the thin solid, dotted, and dashed lines indicate
the first, second, and third terms, respectively, in each of the lines of (5.1)) 
and Ohmic dissipations (thick dotted shaded lines) 
in the cases
$P=0.1$, $\tau=10^5$, $R=6\x10^6$, $P_m=0.15$ (left column),
$P=0.1$, $\tau=10^5$, $R=3\x10^6$, $P_m=3$ (middle column),
$P=200$, $\tau=5\x10^3$, $R=10^6$, $P_m=80$ (right column).
} 
\label{f.16}
\end{figure}

It is of interest to compare the various interactions between velocity
and magnetic field components which sustain dynamo action against
Ohmic dissipation. Zhang and Busse (1989) have listed a total of 31
interaction integrals which contribute to the energy balances obtained
from equations \eqref{induction} for the mean and fluctuating components of
the poloidal and toroidal magnetic fields. Fortunately only a few of
the 31 terms contribute significantly and we list here only the most
important ones, 
\begin{subequations}
\begin{align}
&
\bar p_1 \equiv ( \check v \check g \bar h ), \qquad \bar p_2 \equiv
( \check w \check h \bar h) , \qquad \bar p_3 \equiv ( \check v \check h \bar h ),  \\
&
\bar t_1 \equiv ( \bar w \bar h \bar g),\qquad \bar t_2 \equiv (
\check w \check h \bar g),\qquad \bar t_3 \equiv ( \check v \check g
\bar g), \\
&
\check p_1\equiv( \check v \bar g \check h ) ,\qquad \check p_2\equiv
( \check v \check g \check h ), \qquad \check p_3\equiv( \check v \check h
\check h), \\
&
\check t_1\equiv( \check w \bar g \check g ) ,\qquad 
\check t_2\equiv( \check w \check g \check g ) ,\qquad \check t_3\equiv
( \check w \bar h \check g ),
\end{align}
\end{subequations}
where the first two letters inside the brackets indicate which of the 
interactions between velocity and magnetic field components on the right
hand sides of equations \eqref{induction} counteract the Ohmic dissipation
of the magnetic field component indicated by the last letter inside the
brackets. In the case of chaotic dynamos these integrals tend to
fluctuate wildly and may change their signs. The particular ones
listed above have been plotted as a function of time in 
figure \ref{f.16} together with the corresponding average Ohmic dissipation
density for three representative cases. It is remarkable that the
differential rotation contributes most of the sustenance of the mean
toroidal field not only in the case of low values of $P$ and $P_m$,
but also in the high Prandtl number case. The mean thermal wind is
obviously sufficient to generate the mean toroidal field through the
distortion and stretching of the mean poloidal field. This
$\omega$-effect is thus usually found to dominate the dynamo process
in the case of the zonal field. Only in particularly chaotic dynamos
characterized by a rather weak mean poloidal field is the
$\omega$-effect not effective as in the case $P = 0.1, P_m = 3$ of
figure \ref{f.16}. It is remarkable that the first and second terms of
definitions (5.1a) are strongly anticorrelated in the low Prandtl
number cases of figure \ref{f.16}. Because of their near cancellations the
axisymmetric poloidal field is rather weak in low $P$ dynamos.

The $\alpha$-effect in which the fluctuating magnetic field is
generated through the interaction between fluctuating velocity field
and mean magnetic field plays a far less important role than the
$\omega$-effect. The fluctuating components of the magnetic field are
usually generated through interactions of fluctuating parts of
velocity and magnetic fields except in the case of  high Prandtl
numbers where the mean poloidal field typically dominates.
\begin{figure} \vspace*{4mm}
\mypsfrag{aa01}{$10^{-1}$}
\mypsfrag{aa0}{$10^0$}
\mypsfrag{aa1}{$10^1$}
\mypsfrag{Ex}{\hspace{-0.9cm}$E_x$, $M_x$}
\mypsfrag{1aa5}{\hspace{2mm}$10^5$}
\mypsfrag{1aa0}{\hspace{2mm}$10^0$}
\mypsfrag{1aa1}{\hspace{2mm}$10^1$}
\mypsfrag{1aa2}{\hspace{2mm}$10^2$}
\mypsfrag{1aa3}{\hspace{2mm}$10^3$}
\mypsfrag{1aa4}{\hspace{2mm}$10^4$}
\mypsfrag{1aa5}{\hspace{2mm}$10^5$}
\mypsfrag{2aa1}{$10^1$}
\mypsfrag{2aa2}{$10^2$}
\mypsfrag{2aa3}{$10^3$}
\mypsfrag{2aa4}{}
\mypsfrag{2aa5}{$10^5$}
\mypsfrag{2aa6}{$10^6$}
\mypsfrag{7.5}{7.5} 
\mypsfrag{10} {10}  
\mypsfrag{3}  {3}   
\mypsfrag{1} {1}    
\mypsfrag{c2}{\hspace{-6mm}$P=1$}
\mypsfrag{f1}{\hspace{-6mm}$P=3$}
\mypsfrag{f3}{\hspace{-6mm}$P=5$}
\mypsfrag{f5}{\hspace{-6mm}$P=15$}
\mypsfrag{d} {\hspace{-6mm}$P=1$}
\mypsfrag{g1}{\hspace{-4mm}$P=3$}
\mypsfrag{g3}{\hspace{-4mm}$P=5$}
\mypsfrag{g5}{\hspace{-5mm}$P=15$}
\mypsfrag{Vxxx}{$V_x$, $O_x$}
\mypsfrag{R}{}
\mypsfrag{RR}{\hspace{0.7cm}$P_m$}
\begin{center}
\hspace{10mm}
\begin{tabular}{@{}c@{\extracolsep{1cm}}c}
\epsfig{file=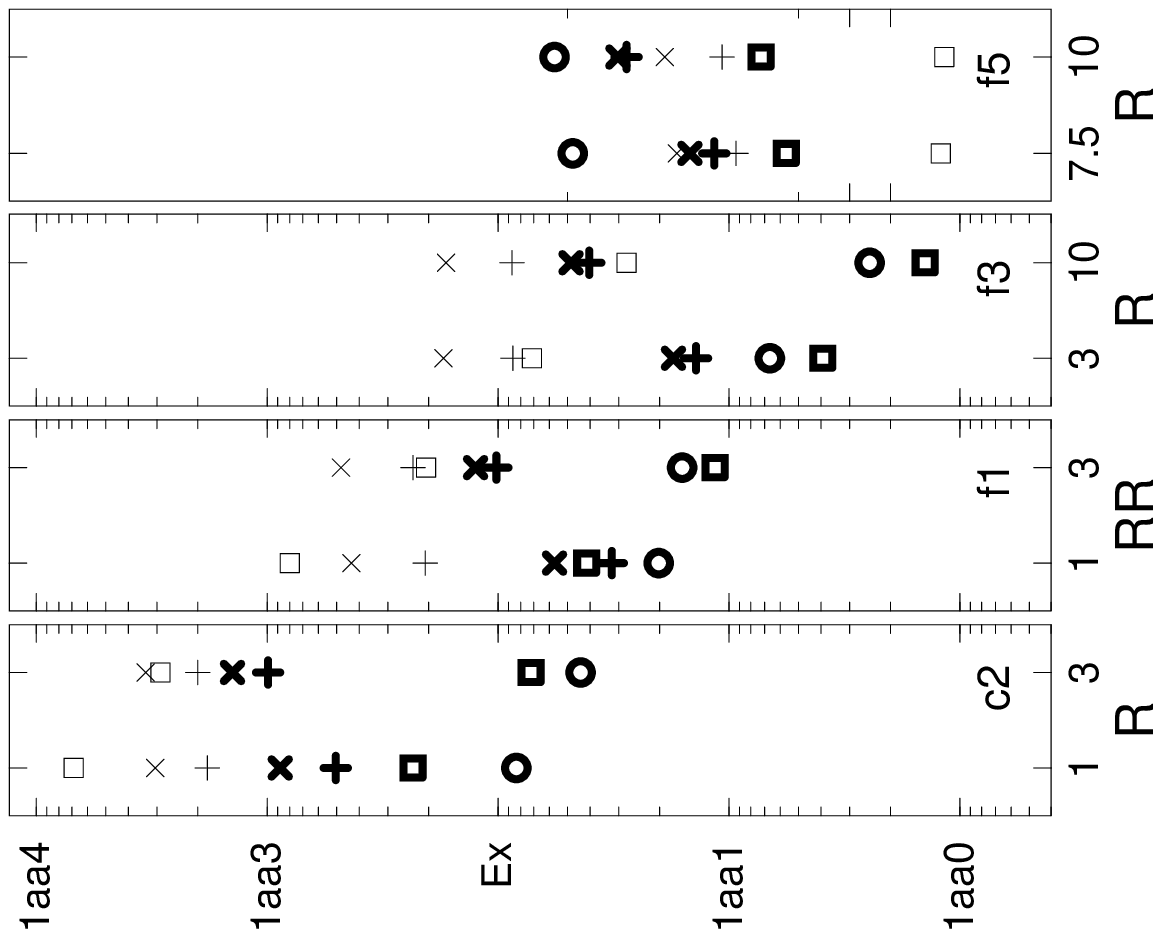,width=8cm,height=6cm,angle=-90} &  
\epsfig{file=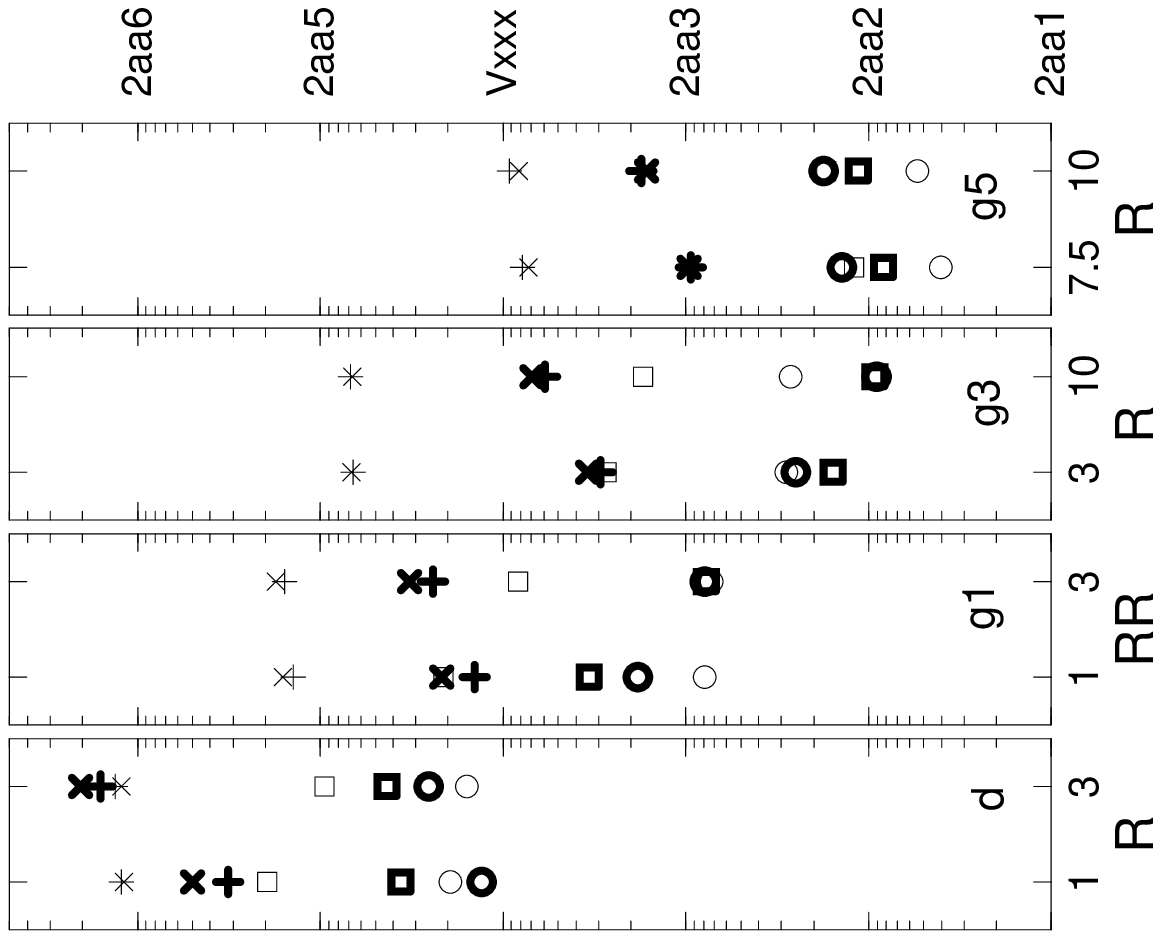,width=8cm,height=6cm,angle=-90}  \\
\end{tabular} 
\end{center}
\caption[]{Kinetic $E_x$ and magnetic $M_x$ energy densities (left)
and viscous $V_x$ and Ohmic $O_x$ dissipations (right) as 
functions of $P_m$ for convection driven dynamos 
for $\tau=5\x10^3$, $R=10^6$ and Prandtl  number
as indicated in the boxes. The highest values of the Elsasser number
for the cases $P=1$, 3, 5, 15 are  $\Lambda=3.02$, $0.31$, $0.37$ and
$0.49$, respectively. The components $\overline{X}_p$, $\overline{X}_t$,
$\check{X}_p$, $\check{X}_t$ (where $X = E$, $M$, $V$, $O$) are
represented by circles, squares, plus-signs and 
crosses, respectively. Kinetic energy densities and viscous
dissipations are shown with light symbols, magnetic energy
densities and Ohmic dissipations are shown with heavy symbols.} 
\label{f.13}
\end{figure}

 One of the least understood questions of convection driven dynamos
is the equilibration of magnetic energy. While in the astrophysical
context an equipartition between kinetic and magnetic energy is 
often favored, an Elsasser number $\Lambda$ of the order unity is
regarded by geophysicists as a good estimate of the magnetic energy
generated by dynamos in planetary cores. In the following we 
shall use the definition 
\begin{equation}
\Lambda= \frac{2 M P_m}{\tau}
\label{e.1000}
\end{equation}
for the Elsasser number where $M$ denotes the magnetic energy density
averaged over the fluid shell and in time, \ie
$M=\overline{M}_p+\overline{M}_t+\check{M}_p+\check{M}_t$. Already
Chandrasekhar (1961)
established that for $\Lambda =1$ a minimum is found of the critical
Rayleigh number for onset of convection in a horizontal layer heated
from below, penetrated by a homogeneous vertical magnetic field and
rotating about a vertical axis.
\begin{figure} \vspace*{4mm}
\mypsfrag{aa01}{$10^{-1}$}
\mypsfrag{aa0}{$10^0$}
\mypsfrag{aa1}{$10^1$}
\mypsfrag{Ex}{\hspace{-0.5cm}$E_x$, $M_x$}
\mypsfrag{Vx}{\hspace{-0.5cm}$V_x$, $O_x$}
\mypsfrag{a0}{\hspace{-1mm}$10^0$}
\mypsfrag{a1}{\hspace{-1mm}$10^1$}
\mypsfrag{a2}{\hspace{-1mm}$10^2$}
\mypsfrag{a3}{\hspace{-1mm}$10^3$}
\mypsfrag{a4}{\hspace{-1mm}$10^4$}
\mypsfrag{a5}{\hspace{-1mm}$10^5$}
\mypsfrag{a6}{\hspace{-1mm}$10^6$}
\mypsfrag{a7}{\hspace{-1mm}$10^4$}
\mypsfrag{a8}{\hspace{-1mm}$10^5$}
\mypsfrag{a9}{\hspace{-1mm}$10^6$}
\mypsfrag{8}{8} 
\mypsfrag{8.5}{8.5} 
\mypsfrag{12} {12}  
\mypsfrag{15} {15}  
\mypsfrag{20} {20}  
\mypsfrag{23} {23}  
\mypsfrag{10} {10}  
\mypsfrag{14} {14}  
\mypsfrag{20} {20}  
\mypsfrag{30}  {30}   
\mypsfrag{50}  {50}   
\mypsfrag{6}  {6}   
\mypsfrag{13} {13}  
\mypsfrag{3}  {3}   
\mypsfrag{5}  {5}   
\mypsfrag{7} {7}    
\mypsfrag{1} {10}    
\mypsfrag{2} {2}    
\mypsfrag{60}  {60}   
\mypsfrag{80}  {80}   
\mypsfrag{k1}{\hspace{-1mm}{\sl (a)}}
\mypsfrag{k2}{\hspace{-1mm}{\sl (b)}}
\mypsfrag{h2}{\hspace{-1mm}{\sl (c)}}
\mypsfrag{d2}{\hspace{-1mm}{\sl (d)}}
\mypsfrag{d7}{\hspace{-1mm}{\sl (e)}}
\mypsfrag{j2}{\hspace{-1mm}{\sl (f)}}
\mypsfrag{j1}{\hspace{-1mm}{\sl (g)}}
\mypsfrag{Exx}{$E_x$, $M_x$}
\mypsfrag{Vxx}{$V_x$, $O_x$}
\mypsfrag{R}{}
\mypsfrag{RRRR}{\hspace{-1cm}$R\x10^{-5}$}
\begin{center}
\hspace{-5mm}
\epsfig{file=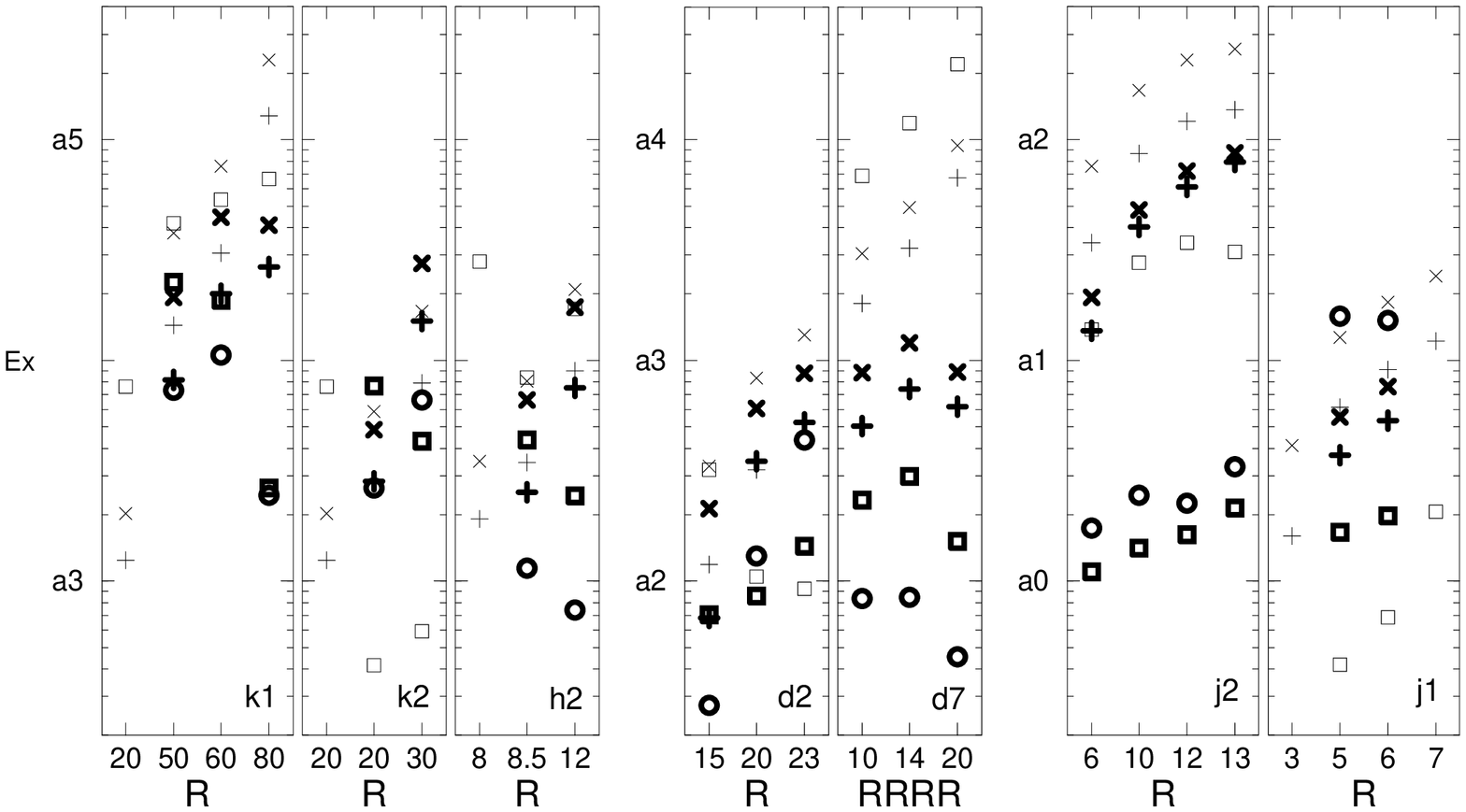,width=8cm,height=32pc,angle=-90} \\[10pt]
\hspace{-5mm}
\epsfig{file=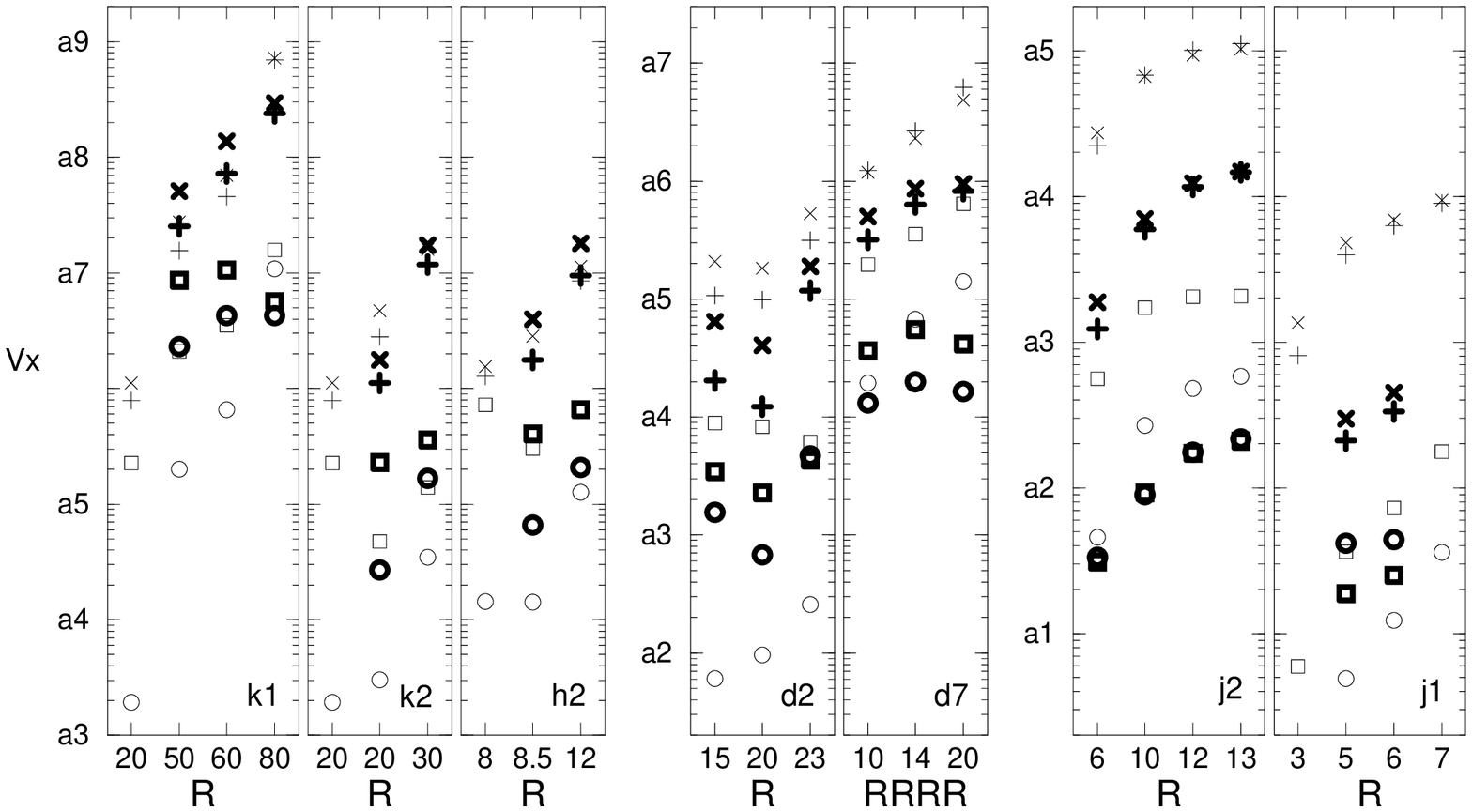,width=8cm,height=32pc,angle=-90} 
\end{center}
\caption[]{Kinetic $E_x$ and magnetic $M_x$ energy densities (upper row)
and viscous $V_x$ and Ohmic $O_x$ dissipations  (lower row) as 
functions of $R$ for convection driven dynamos in the cases 
{\sl (a)} $P=0.1$, $\tau=10^5$, $P_m=0.2$,  $\Lambda= 0.38$; 
{\sl (b)} $P=0.1$, $\tau=10^5$, $P_m=3$,    $\Lambda= 3.22$; 
{\sl (c)} $P=0.1$, $\tau=3\x10^4$, $P_m=1$, $\Lambda= 1.87$; 
{\sl (d)} $P=1$, $\tau=3\x10^4$, $P_m=2$,   $\Lambda= 0.26$; 
(e) $P=1$, $\tau=5\x10^3$, $P_m=1$, $\Lambda= 0.93$; 
(f) $P=5$, $\tau=5\x10^3$, $P_m=10$,  $\Lambda= 0.68$ and
(g) $P=10$, $\tau=5\x10^3$, $P_m=10$, $\Lambda= 0.12$.
The components $\overline{X}_p$, $\overline{X}_t$,
$\check{X}_p$, $\check{X}_t$ (where $X = E$, $M$, $V$, $O$) are
represented by circles, squares, plus-signs and 
crosses, respectively. Kinetic energy densities and viscous
dissipations are shown with light symbols, magnetic energy
densities and Ohmic dissipations are shown with heavy symbols. In case
{\sl (b)} results for a convection driven dynamo as well as for convection
without magnetic field have been plotted in the case $R=2\x10^6$. The
values of $\Lambda$ represent the highest values for each of the boxes.}
\label{f.14}
\end{figure}

In figures \ref{f.13} and \ref{f.14} averaged kinetic and magnetic
energy densities and viscous and ohmic dissipation densities have been
plotted for typical parameter sets of the problem. The Elsasser number
has been computed for the largest magnetic energy $M$ in each of the
boxes and is given in the figure captions. It is evident that
$\Lambda$ rarely exceeds unity and when it does a physical reason of a
high value of $\Lambda$ is not obvious. In particular, in the cases of
figure \ref{f.13} where only $P_m$ is varied it is hard to understand
why the Elsasser number increases more strongly than $P_m$
itself. Another way of looking at the problem is to determine the
highest value of $\Lambda$ as a function of the Rayleigh number. A
systematic study of the variation of magnetic energies with increasing
$R$ has been undertaken by Grote and Busse (2001) in the case $P=P_m=1$,
$\tau=5\x10^3$. This study has been extended since that time in order
to obtain reliable time averages, see figure 11 of Busse and Simitev
(2004b). Here the   case of maximal $M$ is attained for $R=1.4\x 10^6$ in part
{\sl (e)} of figure \ref{f.14} and corresponds to a value of about
unity for $\Lambda$. But this result may be accidental as is suggested by the
magnetic fields in the cases of dynamos for low values of $P$ and
$P_m$. Here again we find in the case  {\sl (a)} of figure \ref{f.14}
that magnetic energy has reached a value close to its maximum as a
function of $R$, but the corresponding value $0.38$ of $\Lambda$ is
significantly lower. A much higher value of $\Lambda$ is obtained,
on the other hand, when $P_m=0.2$ 
is replaced by $P_m=3$ even though $M$ may still increase with a
further increase in $R$.  We thus conclude that the criterion
$\Lambda\approx 1$ for the equilibration of the magnetic energy is
useful as a rough estimate, but the dependence of the equilibrium
magnetic energy on $P_m$ is actually weaker than suggested by the
definition \eqref{e.1000} of $\Lambda$.

There are a number of other features exhibited by the diagrams of
figures \ref{f.13} and \ref{f.14}. While the energy densities of the
fluctuating components of the magnetic field usually exceed those of
the mean components this situation is reversed at sufficiently high
Prandtl numbers for the poloidal part of the field as is evident from
the last columns of figure \ref{f.13} as well as of figure
\ref{f.14}. This changeover occurs rather suddenly at about $P=8$ for
$\tau=5\x10^3$ as can be noticed
in the  comparison of cases {\sl (f)} and {\sl (g)} of figure
\ref{f.14}.  A closer inspection shows that the replacement of the
geostrophic differential rotation by the thermal wind and the
accompanying growth of the polar zonal flux tubes is responsible for
this change in the dynamo process. It should be noticed that this
changeover is not evident in the plots of the dissipation densities
since the latter are always much lower for the mean components than
for the fluctuating components. 

Another property that changes with increasing Prandtl number  is the
fact that for $P$ of the order unity or less the energy of the mean
toroidal field exceeds that of the mean poloidal field except for some
high Rayleigh number cases where polar convection begins to
dominate. 
\begin{figure}
\vspace*{4mm}
\begin{center}
\epsfig{file=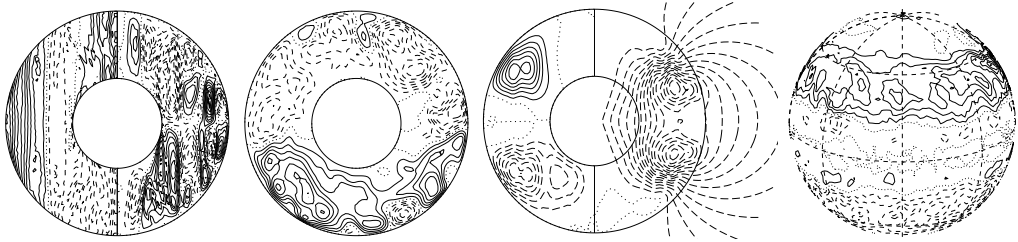,width=32pc,angle=0,clip=} 
\end{center}
\caption{Convection driven dynamo for $P=0.01$,
  $\tau=10^5$, $R=6\x 10^5$ and $P_m=0.5$. The first plot shows lines of constant
  $\overline{u}_\varphi$ in the left half and streamlines, $r \sin \theta
  \dd_\theta \overline{v}=$ const.~in the right half. The second plot
  shows streamlines in the equatorial plane, $r \dd_\varphi v=$
  const.. The left half of the third plot shows lines of constant
  zonal magnetic field $\overline{B}_\varphi$ and the right
  half shows  meridional field lines, $r\sin\theta \dd_\theta
  \overline{h} =$ const.. The last plot shows 
 lines of constant $B_r$ at the surface $r=r_o+0.5$.}
\label{f.15a}
\end{figure}
For dipolar dynamos obtained for $P=0.01$, $\tau=10^5$ with Rayleigh
numbers of the order $10^6$ and $0.2\leq P_m \leq 0.5$ the energy of
the mean toroidal field exceeds that of the mean poloidal field by
more than a factor $10^2$. For example, in the case $R=10^6, P_m=0.5$
the time averaged magnetic energy densities $\overline{M}_p=146$ and
$\overline{M}_t=2.26 \x 10^4$ are found. The structure of velocity and
magnetic fields in this case is apparent from figure 15. The
axisymmetric components of the magnetic field are almost steady in
time. But the differential rotation in the polar regions may change
its sign. At the particular time of figure 15 the zonal flows near the
north- and south-poles have opposite signs. The fluctuating component
of convection is dominated by the $m=1$-mode and exhibits the
attachment to the equatorial boundary as must be expected for
inertial convection according to the discussion in section 3. 
For Prandtl numbers of the
order $5$ and higher, on the other hand, the energy of the mean
poloidal field always exceeds that of the mean toroidal field, since 
the differential rotation which creates the latter is not strong
enough. But this changeover is rather gradual and depends more on the
Rayleigh number than the other changeover discussed above.
\begin{figure} \vspace*{4mm}
\mypsfrag{aaa1}{$2\x10^5$}
\mypsfrag{aaa}{$10^6$}
\mypsfrag{aa1}{$10^6$}
\mypsfrag{b1}{$10^0$}
\mypsfrag{1}{$1$}
\mypsfrag{10}{$10$}
\mypsfrag{b2}{$10^1$}
\mypsfrag{b3}{$10^2$}
\mypsfrag{b4}{$10^3$}
\mypsfrag{b5}{$10^4$}
\mypsfrag{R}{$R$}
\mypsfrag{Emt}{$\overline{E}_t$}
\mypsfrag{Eft}{$\check{E}_t$}
\mypsfrag{Nui}{$Nu_i$}
%
\begin{center}
\hspace*{0mm}
\epsfig{file=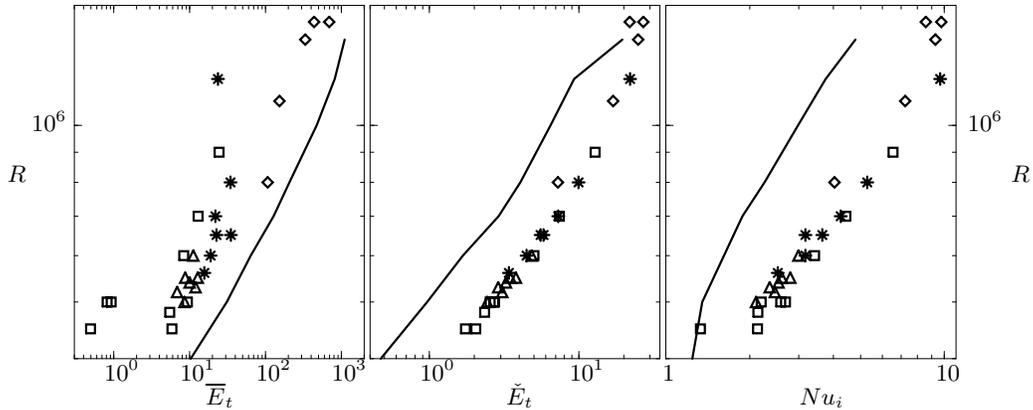,height=32pc,angle=-90}
\end{center}
\caption[]{Time-averaged kinetic energy densities and Nusselt number
  $Nu_i$ of non-magnetic 
  convection (thick lines) and of quadrupolar (diamonds), mixed (stars),
  hemispherical (triangles) and chaotic dipolar dynamos
  (squares). Both non-magnetic 
  convection and dynamo solutions are in the case $P=1$ and $\tau=10^4$. 
  Values of $P_m$ of the dynamo cases decrease from $10$ to $0.4$ as
  $R$ increases. The densities $\overline{E}_t$,
  $\check{E}_t$ and $Nu_i$ are shown in the left, middle and right panel,
  respectively.}
\label{f.15}
\end{figure}

Since in some of the cases of figure \ref{f.14} the kinetic energies
of convection in the absence of a magnetic field are shown, the effect
of the magnetic field on convection becomes evident. In particular in
the cases with $P$ of the order unity or less the energy of the mean
toroidal velocity field is strongly reduced by the dynamo action in that the
Lorentz force leads to a braking of the differential rotation as has
been discussed by Grote and Busse (2001). On the other hand the
fluctuating components of convection are enhanced since the magnetic
field prevents the onset of relaxation oscillations in which
convection occurs only intermittently as has been discussed in
connection with figure \ref{f.02}. This effect is clearly demonstrated
in case {\sl (b)} of figure \ref{f.14} where the results for
convection with and without magnetic field are compared for
$R=2\x10^6$. The same features are also revealed in
figure \ref{f.15} where a more general comparison between  convection with
and without magnetic fields is shown. This figure also demonstrates
that the symmetry of the magnetic field does not play a major role in
the considerations of  energetic aspects of dynamos. 

For Prandtl
numbers larger than unity the effect of the magnetic field is much
reduced which is evident from the fact that the increasing magnetic
energy for increasing $P_m$ does not change the energy of the
fluctuating components of convection as can be seen in figure
\ref{f.13}. Only the differential rotation is reduced through the
strengthening of the Lorentz force. The diminished influence of
magnetic field on convection is also evident from the property that 
typically viscous dissipation exceeds Ohmic dissipation by far for
values of $P$ of the order $5$ or higher, while the two sinks of energy
are more comparable at low values of $P$.

\section{Validity of the magnetostrophic approximation}

\begin{figure} \vspace*{4mm}
\mypsfrag{Mmp}{$\overline{M}_p P$}
\mypsfrag{Mmt}{$\overline{M}_t P$}
\mypsfrag{Mfp}{$\check{M}_p P$}
\mypsfrag{Mft}{$\check{M}_t P$}
\mypsfrag{a}{\sl(a)}
\mypsfrag{b}{\sl(b)}
\mypsfrag{c}{\sl(c)}
\mypsfrag{d}{\sl(d)}
\mypsfrag{e}{\sl(e)}
\mypsfrag{f}{\sl(f)}
\mypsfrag{1}{1}
\mypsfrag{10}{10}
\mypsfrag{100}{100}
\mypsfrag{P}{$P$}
\mypsfrag{a01}{\hspace*{-2mm}$10^{-1}$}
\mypsfrag{aa0}{$10^0$}
\mypsfrag{aa1}{$10^1$}
\mypsfrag{aa2}{$10^2$}
\mypsfrag{aa3}{$10^3$}
\mypsfrag{aa4}{$10^4$}
\mypsfrag{aa5}{$10^5$}
%
\epsfig{file=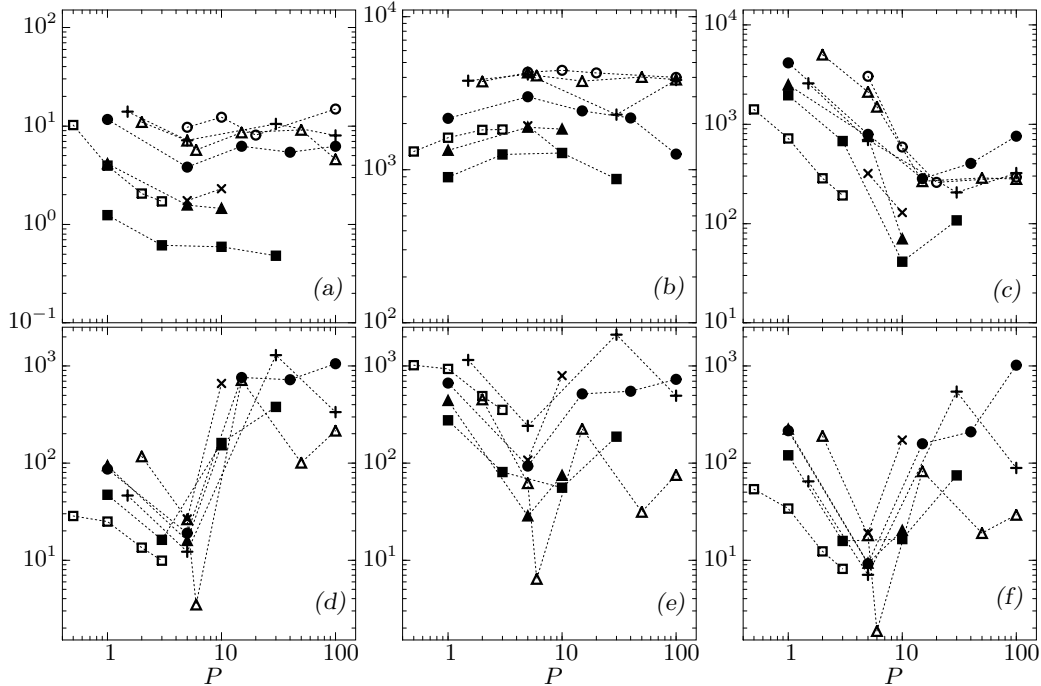,height=32pc,angle=-90}
\caption[]{
Kinetic energy densities $\overline{E}_p$ (in {\sl (a)}), 
$\check{E}_t$ (in {\sl (b)}) and $\overline{E}_t$ (in {\sl (c)}), all
multiplied by $P^2$ as functions of  $P$ in the case
$\tau=5 \x 10^3$. The dynamos corresponding to 
$\kappa/\lambda=1$, $R=5\x10^5$ are  indicated by solid squares,
    to $\kappa/\lambda=1$, $R=6\x10^5$ by solid triangles,
    to $\kappa/\lambda=1$, $R=8\x10^5$ by solid circles,  
    to $\kappa/\lambda=2$, $R=6\x10^5$ by crosses,
    to $\kappa/\lambda=2$, $R=10^6$ by plus-signs,
    to $\kappa/\lambda=5$, $R=6\x10^5$ by empty squares,
    to $\kappa/\lambda=0.5$, $R=10^6$ by empty triangles and
    to non-magnetic convection with $R=10^6$ by empty circles.
The second row shows the corresponding magnetic energy densities
    multiplied by $P$. }
\label{f.12}
\end{figure}

Because convection driven dynamos in rotating systems depend on
a rather large number of parameters it is desirable to
eliminate one or more parameters through reductions of the basic
equations. Among the nonlinear advection terms the momentum advection
term appears to be most expendable since it does not seem to be
essential for convection driven dynamos. In this way the 
magnetostrophic approximation is obtained in which the acceleration of fluid
particles is neglected in comparison to the Coriolis force and the Lorentz
force. In general this approximation can be easily obtained through the
replacement of the viscous time scale $d^2/\nu$ by the time scale
$d^2/(\kappa^{1-\gamma} \lambda^{\gamma})$ with $0 \leq \gamma \leq 1$
which is intermediate between the thermal and magnetic diffusion time
scales. When $\sqrt{\rho \mu \kappa^{1-\gamma} \lambda^\gamma \nu}/d$
is used as scale of the magnetic 
field the basic dimensionless equations of motion, equation of
induction and the heat equation can be written in the form
\begin{subequations}
\label{magn}
\begin{align}
\label{magn1}
&\hspace*{-1.5cm}
(P^{1-\gamma} P_m^\gamma)^{-1} (\partial_t \vec u + \vec u \cdot
\nabla \vec u ) + \tau \bec k \times \bec u = - \nabla \pi + \Theta
\bec r + \nabla^2 \bec u + ( \nabla \times \bec B) \times \bec B\\  
\label{magn2}
&\hspace*{-1.5cm}
(P_m/P)^\gamma \left( \partial_t \vec B + \vec u \cdot \nabla \vec B
  - \bec B \cdot \nabla \bec u \right) = \nabla^2 \bec B \\
\label{magn3}
&\hspace*{-1.5cm}
(P/P_m)^{1-\gamma} \left(\partial_t \Theta + \vec u \cdot \nabla
  \Theta \right) = R \bec u \cdot \bec r + \nabla^2 \Theta 
\end{align}
\end{subequations}
where $\vec k$ is the unit vector parallel to the axis of
rotation. From the form of equations \eqref{magn} it is clear that the
magnetostrophic approximation should certainly be valid in the limit
$P^{1-\gamma} P_m^\gamma \longrightarrow \infty$. 
In the following we shall focus on the case $\gamma = 0$. In figure
\ref{f.12} the energy densities have been plotted for
fixed values of $R, \tau$ and of $\kappa/\lambda$. Since the
fluctuating poloidal energy density $\check E_p$ always amounts to about 50
percent of the corresponding toroidal one it has not been plotted. It
can be seen that the kinetic energies tend to become independent of \P
with increasing \P in 
accordance with the magnetostrophic assumption. The energy density $\overline{E}_t$
representing the differential rotation is the only exception as
expected. Little indication of an approach towards the validity of the
magnetostrophic approximation is found, however, when the magnetic
energy densities are considered. As has already been mentioned the
dynamo process is rather sensitive to the presence of the differential
rotation and much higher values of \P may be needed before the
magnetostrophic regime is approached. It is remarkable to see the
distinct minimum of magnetic energies near $P = 8$ which corresponds
to change in the structure of the magnetic field as has been
mentioned. The transition from a geostrophic differential rotation to a
thermal wind one appears to be mainly responsible for this feature. 

Results obtained on the basis of the magnetostrophic approximation in
the case $\gamma = 0$ 
are by definition independent of $P$. In particular the ratio between
the magnetic energy and kinetic energy will be proportional to \P
(Glatzmaier and Roberts, 1995)  as is evident from the different
scales used for the velocity and for $\vec B/ \sqrt{\rho \mu}$ in
equation \eqref{magn1}. The ratio between Ohmic and viscous dissipation
would be independent of \P and would depend only on
$\kappa/\lambda$. This latter parameter seems to be even more
important than the magnetic Prandtl number for convection driven
dynamos. At least for Prandtl number of the order unity or less the condition
$1\lesssim P_m/P$  appears to apply for the minimum value $P_m$ for
which dynamo action can be obtained as is indicated by a comparison of
dynamos obtained for $P=1$ (see figure 1 of  Grote \etal,
2002) and for $P=0.1$ (see figure \ref{f.09}).   

Similar conclusions as have been reached for $\gamma = 0$ may be
reached for $\gamma = 1$. In this case the energy densities of figure
\ref{f.12} can be plotted as a function of $P_m$. For the cases
$\kappa/\lambda = 1$ the plots remain unchanged with $P_m$ replacing
$P$ at the abscissa. For other values of $\kappa/\lambda$ the
respective data would just be shifted. Since no particular new insight
appears to be evident we have not included such a figure. The general
conclusion to be drawn is that the magnetostrophic approximation seems
to become valid only for fairly high values of the parameter
$P^{1-\gamma} P_m^\gamma$ with $0 \leq \gamma \leq 1$. It should also
be remembered that properties such as inertial convection and
geostrophic differential rotation generated by Reynolds stresses
disappear when the magnetostrophic approximation is used. Kuang and
Bloxham (1996) have introduced for this reason a compromise in that
the magnetostrophic approximation is applied only to the
non-axisymmetric part of the equations of motion. But their main
result that the ratio of magnetic to kinetic energy reaches $10^3$ for
$P = 1$ is far removed from properties of other convection driven
spherical dynamos described in the literature. 

\section{Effects of boundary conditions on dynamo solutions}

\begin{figure} \vspace*{4mm}
\begin{center}
\epsfig{file=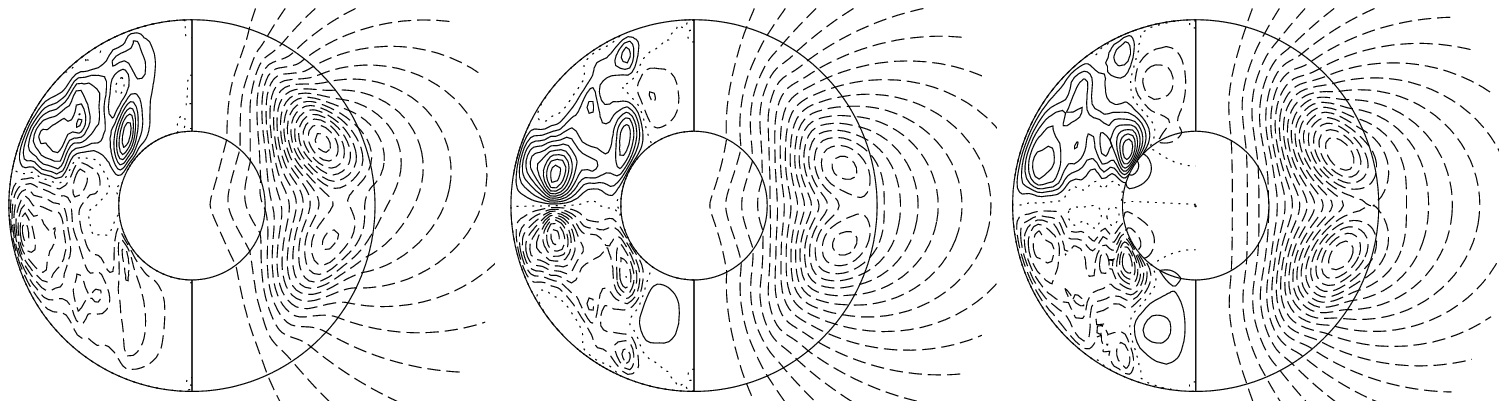,width=28pc}
\end{center}
\vspace*{2mm}
\caption[]{Effects of various boundary conditions on dynamo
  solutions for $P=0.1$, $\tau=10^5$, $R=4\x10^6$, $P_m=0.5$ in the dipolar cases {\bf A}, {\bf B}, {\bf C}  (from left to right) of table \ref{t.8.10}.
  The left halves of each plot show lines of constant zonal
  magnetic field $\overline{B}_\varphi$ and the right
  halves show  meridional field lines, $r\sin\theta \dd_\theta \overline{h} =$ const., at particular moments in time.}
\label{df.8.40}
\vspace*{4mm}
\begin{center}
\epsfig{file=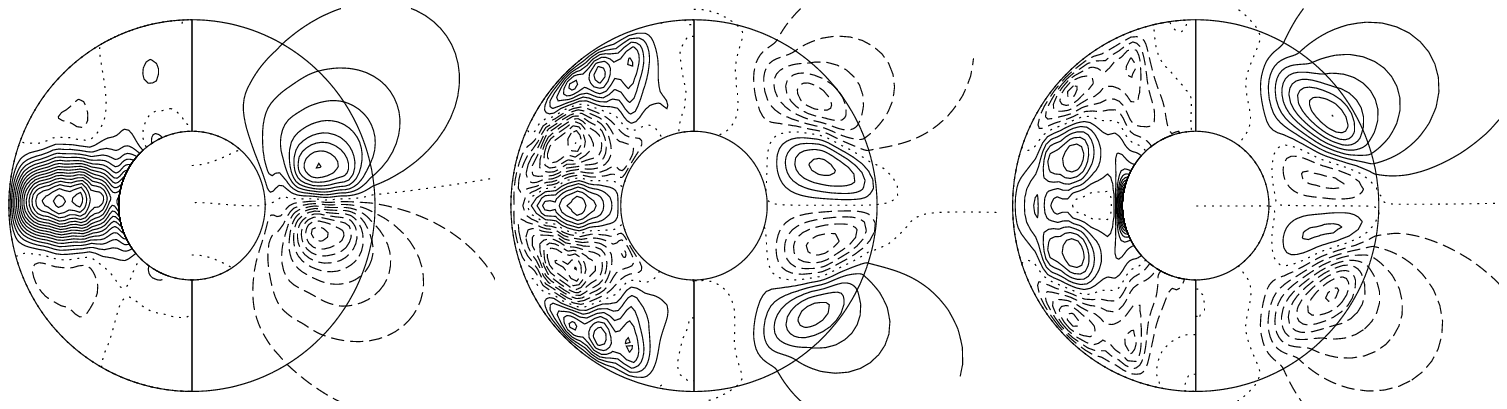,width=28pc}
\end{center}
\vspace*{2mm}
\caption[]{Same as figure \ref{df.8.40} but for the quadrupolar cases {\bf D} and {\bf A} and
  {\bf D} for the parameter values $P=5$, $\tau=5\x10^3$,
  $R=8\x10^5$, $P_m=3$.}
\label{df.8.50}
\end{figure}

Because published solutions of convection driven dynamos in rotating
spherical shells are based on a variety of parameter values
and satisfy different mechanical, thermal and 
magnetic boundary conditions, a systematic comparison between
various models is difficult to conduct. An attempt to
investigate the differences between  numerical models based
on various assumptions has most recently been made by Kutzner and
Christensen (2000, 2002) who consider
thermal and compositional driving with internal or external
distribution of energy sources and a variety of thermal boundary
conditions. Here we complement their studies with a few remarks
on the possible effects of various velocity and 
magnetic boundary conditions. For this purpose we
have selected two typical dynamo solutions. The
dynamo at parameter values $P=0.1$, $\tau=10^5$, $R=4\x10^6$, $P_m=0.5$ has a
dipolar symmetry while the solution at $P=5$, $\tau=5\x10^3$,
$R=8\x10^5$, $P_m=3$ has a quadrupolar one. The results for the
standard boundary conditions \eqref{vbc} and \eqref{mbc} used in this
paper are listed under {\bf  A}. 
\begin{table}
%
%
\caption{Time-averaged global properties of dynamos with various
    velocity and magnetic boundary conditions as follows. {\bf A}:
    stress-free and insulating, {\bf B}: no-slip and insulating, 
    {\bf C}: no-slip and a finitely-conducting inner core and  {\bf D}:
    stress-free and a  perfectly-conducting inner core. The
    predominant symmetry type 
    is indicated with ``D'' if dipolar, ``Q'' if quadrupolar and ``--'' if
    the dynamo is decaying. In the case {\bf B} in left part of the
    table results for both a dynamo solution and a non-magnetic
    convection case are given.} 
\vspace{4mm}
\begin{center}
\scriptsize
\begin{tabular}{@{}||c@{\ }|ccccc|cccc||}\hhline{|t:==========:t|} 
 & 
\multicolumn{5}{c|}{$P=0.1$, $\tau=10^5$, $R=4\x10^6$, $P_m=0.5$} &
\multicolumn{4}{c||}{$P=5$, $\tau=5\x10^3$, $R=8\x10^5$, $P_m=3$ \rule{0in}{3ex}} \\[1ex] 
 & {\bf A} & {\bf B} & {\bf B} & {\bf C} & {\bf D} & {\bf A} & {\bf  B} &{\bf  C} &{\bf D}  \\[1ex]  \hhline{||----------||} 
 Type & D & D & -- & D & Q & Q & -- & -- & Q \rule{0in}{3ex}\\
& & & & & & & & &  \\
  $\overline{E}_p $  & .286$\x10^2$ & .112$\x10^2$ & .939$\x10^2$ &.108$\x10^2$  &.299$\x10^2$ & .157 &        .611         & .523         & .147 \\
  $\overline{E}_t $  & .599$\x10^4$ & .647$\x10^3$ & .757$\x10^5$ & .807$\x10^3$ &.764$\x10^4$ & .533$\x10^2$ & .105$\x10^2$ & .106$\x10^2$ & .528$\x10^2$ \\
  $\check{E}_p    $  & .142$\x10^5$ & .115$\x10^5$ & .221$\x10^5$ &.121$\x10^5$ &.138$\x10^5$ & .574$\x10^2$ & .614$\x10^2$ & .611$\x10^2$ & .566$\x10^2$ \\
  $\check{E}_t $     & .336$\x10^5$ & .257$\x10^5$ & .540$\x10^5$ &.273$\x10^5$ &.247$\x10^5$ & .119$\x10^3$ & .983$\x10^2$ & .952$\x10^2$ & .119$\x10^3$ \\
& & & &  & & & & &  \\
  $\overline{M}_p $  & .129$\x10^5$ & .392$\x10^5$ & -- & .355$\x10^5$  &.768$\x10^3$ & .487$\x10$  & -- & -- & .466$\x10$ \\
  $\overline{M}_t $  & .133$\x10^5$ & .879$\x10^4$ & -- & .716$\x10^4$  &.538$\x10^5$ & .325$\x10$  & -- & -- & .336$\x10$ \\
  $\check{M}_p    $  & .136$\x10^5$ & .169$\x10^5$ & -- & .147$\x10^5$  &.148$\x10^5$ & .818$\x10$  & -- & -- & .837$\x10$ \\
  $\check{M}_t $     & .307$\x10^5$ & .345$\x10^5$ & -- & .327$\x10^5$  &.448$\x10^4$ & .109$\x10^2$& -- & -- & .110$\x10^2$ \\
& & & &  & & & & &  \\
$\overline{V}_p $  & .643$\x10^5$ & .104$\x10^6$ & .107$\x10^8$ & .122$\x10^6$ &.462$\x10^5$ & .151$\x10^3$  & .111$\x10^4$  & .686$\x10^3$ & .140$\x10^3$ \\
$\overline{V}_t $  & .584$\x10^6$ & .366$\x10^6$ & .277$\x10^8$ &.348$\x10^6$ &.709$\x10^6$ & .182$\x10^4$  & .342$\x10^4$  & .245$\x10^4$ & .179$\x10^4$ \\
$\check{V}_p    $  & .133$\x10^8$ & .179$\x10^8$ & .448$\x10^8$ &.191$\x10^8$ &.125$\x10^8$ & .414$\x10^5$  & .524$\x10^5$  & .528$\x10^5$ & .406$\x10^5$ \\
$\check{V}_t $     & .229$\x10^8$ & .354$\x10^8$ & .821$\x10^8$  & .372$\x10^8$ &.171$\x10^8$ & .455$\x10^5$  & .580$\x10^5$  & .568$\x10^5$ & .450$\x10^5$ \\
& & & &  & & & & &  \\                                                                
  $\overline{O}_p $  & .626$\x10^6$ & .129$\x10^7$ & -- & .120$\x10^7$ &.874$\x10^5$ & .544$\x10^3$ & -- & -- & .520$\x10^3$ \\ 
  $\overline{O}_t $  & .148$\x10^7$ & .944$\x10^6$ & -- & .890$\x10^6$ &.222$\x10^7$ & .359$\x10^3$ & -- & -- & .372$\x10^3$ \\ 
  $\check{O}_p    $  & .125$\x10^8$ & .138$\x10^8$ & -- & .133$\x10^8$ &.106$\x10^8$ & .468$\x10^4$ & -- & -- & .488$\x10^4$  \\
  $\check{O}_t $     & .253$\x10^8$ & .268$\x10^8$ & -- & .276$\x10^8$ &.362$\x10^7$ & .600$\x10^4$ & -- & -- & .587$\x10^4$ \\
& & & &  & & & & &  \\   
  $\Lambda $         & .707  & .995  & -- & .902  & .740  & .032 &
  -- & -- & .032\\
  $Rm $            & 164   & 138   & -- &  142  &  152  & 64 & 55 & 55 & 64\\
  $Nu_i$             & 2.125 & 2.275 & 2.442 &  2.316 & 1.563 & 11.285 & 12.968& 12.729 & 11.194\\[1ex] 
\hhline{|b:==========:b|}
\end{tabular}
\end{center}
\label{t.8.10} 
\normalsize
\end{table}   
We have repeated the
simulations of these two cases for three of models with different
combinations of boundary conditions and present the results in table
\ref{t.8.10}. For type {\bf B} no-slip and insulating
conditions have been adopted, for type {\bf C} we have chosen no-slip conditions and
a finitely-conducting inner core and finally type {\bf D} has
stress-free velocity boundary conditions and a  perfectly-conducting
inner core. Thermal conditions have remained unchanged in that
temperatures are fixed at the boundaries in all four
models. Snapshots of the spatial structure of the
magnetic field are shown in figures \ref{df.8.40} and \ref{df.8.50}. 

The replacement of the stress-free condition by the no-slip condition
at the inner and outer boundaries leads to considerable changes in the
structure of the velocity field which are even more dramatic in the
absence of a magnetic field. The differential rotation is strongly
inhibited by the Ekman layers which tend to oppose any deviation from
rigid body rotation. The circulation induced by the Ekman layers, on
the other hand, promotes the heat transport and and is responsible for
the fact that the energy component $\overline{E}_p$ is typically much
larger in the case of no-slip boundaries than in the case of
stress-free ones. But the main effect of the rigid boundaries is the
suppression of the destruction of convection columns by the shearing
action of the differential rotation. Relaxation oscillations which are
predominant for Prandtl numbers of the order unity or less in the
presence of stress-free conditions have not yet been found when
no-slip conditions are used. It must be expected, however, that the
properties of convection with rigid boundaries will approach those of
convection with free boundaries as the Coriolis number $\tau$
increases and the influence of the Ekman layer decreases. 

The effects caused by the differences in the velocity boundary
conditions are reduced in the presence of 
a magnetic field generated by the dynamo process. The differential
rotation is reduced by the action of the Lorentz force for both types
of boundary conditions. But while the energy of the fluctuating
components of convection is amplified by the presence of the magnetic
field this effect usually does not happen in the case of no-slip
boundaries as is evident from case {\bf B} of table \ref{t.8.10}. On
the contrary, the magnetic degrees of freedom compete with kinetic ones
for the same source of energy. Nevertheless, in spite of the lower
kinetic energies in case {\bf B} in comparison to those of case {\bf
A}, case {\bf B} still exhibits a higher convective heat transport as
is evident from the Nusselt number. 

The comparison of convection driven dynamos in case of $P=5$ as
presented by the right side of table \ref{t.8.10} has been less successful
since it was not possible to obtain dynamos in the cases {\bf B} and
{\bf D} even after the Rayleigh number has been  increased up to twice
the indicated value. Except for the axisymmetric components of the
velocities the kinetic energies of the cases {\bf A} and {\bf B} are
surprisingly close which indicates that the effect of no-slip
boundaries on convection is similar to that of a dynamo generated
magnetic field. The closeness of the energy results of  the cases {\bf A}
and {\bf D} for both values of $P$ and of cases {\bf B} and {\bf C}
for $P=0.1$ indicates that the magnetic boundary conditions at the
inner core have rather little influence on the dynamo process. It is
remarkable, however, that in case {\bf D} a quadrupolar dynamo is
realized instead of the dipolar one in the standard case {\bf A}. Even
more remarkable is the property that this quadrupolar dynamo does not
oscillate in time as has been found for all quadrupolar dynamos with
the boundary conditions \eqref{mbc}. From a
general point of view it must be expected that for turbulent dynamos
the inner core can not exert a strong influence on the dynamo process
since the highly fluctuating magnetic field will not penetrate far
into the inner core nor is it likely that strong fluxes will
accumulate there since they are subject to Ohmic decay.  Wicht (2002)
has arrived at similar conclusions.

\section{Concluding remarks}
In this paper the analysis of convection driven dynamos has focused on
the Prandtl number dependence while other parameters such as $R$ and
$\tau$ have not been varied as much, but instead have been kept at
reasonably high values which are numerically accessible with adequate
resolution in space and in time. In the case of the Rayleigh number
the numerical limitations are not felt as strongly as in the case of
$\tau$ since dynamos often seem to disappear or at least exhibit
decreasing energies of the axisymmetric components of their magnetic
fields as $R$ is increased beyond an optimal value. More restrictive
are the limits on the Coriolis number $\tau$ which have prevented so
far the attainment of dynamos dominated by Ohmic dissipation. While
the latter exceeds the viscous dissipation in some of the low Prandtl
number dynamos, sufficient viscous friction always seems to be
required to obtain numerically well-behaved solutions. Ohmic
dissipation also helps in this respect and permits solutions at values
of $R$ and $\tau$ which are not accessible at the same numerical
resolution without magnetic field. But Ohmic dissipation does not
necessarily prevent tangential discontinuities of the velocity field
which are possible in the absence of viscous friction. 

A primary goal of dynamo simulations are results which can be compared
quantitatively with properties of the geomagnetic field and its
variations in time. The geodynamo appears to exhibit features which
are similar to those of convection driven dynamos with large values of
$P$ as well as to those with $P$ of the order of unity. The fact that
the magnetic energy exceeds the kinetic energy in the Earth's core by
a factor of the order $10^3$ together with the property that the
axisymmetric poloidal component of the geomagnetic field dominates
over the non-axisymmetric components suggest that the geodynamo
resembles a high-$P$ dynamo. On the other hand, the strong variations
of the amplitude of the magnetic field on the magnetic diffusion
timescale together with the torsional oscillations which manifest
themselves as ``jerks'' on a much shorter timescale (Bloxham {\it et al.},
2002) indicate a relationship to dynamos with $P \lesssim 1$ (Busse and
Simitev, 2004b). Of course, these similarities may change when more
realistic dynamo simulations with much lower values of the magnetic
Prandtl number $P_m$ become possible.  But they could also indicate
that the two sources of buoyancy apparently present in the Earth's
core, -- the thermal one characterized by $P<1$ and the compositional
one characterized by $P \gg 1$ --, should be distinguished in dynamo
simulations. An indication that new dynamical effects become important
in the presence of buoyancy sources with rather different
diffusivities has been given by the linear analysis of Busse (2002b).


\begin{thebibliography}{}
\bibitem[Ahlers \& Xu (2001)]{AhlersXu2001}
\textsc{Ahlers, G., and Xu, X.} 2001 Prandtl number dependence of heat transport in
turbulent Rayleigh-B\'enard convection. {\it Phys. Rev. Lett.}, {\bf 86},
3320--3323.
\bibitem[Ardes \etal (1997)]{Ardesetal1997}
\textsc{Ardes, M., Busse, F.H. and Wicht, J.} 1997 Thermal convection
in rotating spherical shell. {\it Phys. Earth Planet. Inter.}, {\bf 99},
55--67.
\bibitem[Bloxham \etal (2002)]{Bloxhametal2002}
\textsc{Bloxham, J., Zatman, S., and Dumberry, M.} 2002 The origin of geomagnetic
jerks. {\it Nature} {\bf 240} {65--68}.
\bibitem[Braginsky \& Roberts (1995)]{BraginskyRoberts1995}
\textsc{Braginsky, S. I., and Roberts, P.H. } 1995 Equations governing convection in Earth's core and the geodynamo. {\it Geophys. Astrophys. Fluid Dyn.} {\bf 79} 1--97.
\bibitem[Busse (2002a)]{Busse2002a}
\textsc{Busse, F.H.} 2002a Convection  flows in rapidly rotating spheres and their
 dynamo action. {\it Phys. Fluids} {\bf 14} 1301--1314.
\bibitem[Busse (2002b)]{Busse2002b}
\textsc{Busse, F.H.} 2002b Is low Rayleigh number convection possible in the Earth's
core? {\it Geophys. Res. Letts.} {\bf 29} GLO149597.
\bibitem[Busse \& Cuong (1997)]{BusseCuong1997}
\textsc{Busse, F.H.  and Cuong, P.G.} 1977 Convection in rapidly rotating spherical
fluid shells. {\it Geophys. Astrophys. Fluid Dyn.} {\bf 8} 17--41.
\bibitem[Busse \etal (2003)]{Busse2003}
\textsc{Busse, F.H., Grote, E., and Simitev, R.}  2003 
  Convection in rotating spherical shells and its dynamo
    action, pp. 130-152 {\it  in "Earth's Core and Lower Mantle", C.A.Jones,
  A.M.Soward and K.Zhang, eds., Taylor \& Francis}
\bibitem[Busse \etal (1998)]{Busse1998}
\textsc{Busse, F.H., Grote, E., and Tilgner, A.} 1998 On convection driven dynamos in
rotating spherical shells. {\it Studia geoph. et geod.} {\bf  42},
211--223.
\bibitem[Busse \& Simitev (2004a)]{BusseSimitev2004a}
\textsc{Busse, F.H., and Simitev, R.} 2004a Inertial convection in rotating  fluid
spheres. {\it J. Fluid Mech.} {\bf 498} 23--30.
\bibitem[Busse \& Simitev (2004b)]{BusseSimitev2004b}
\textsc{Busse, F.H., and Simitev, R.} 2004b Convection in rotating spherical fluid
shells and its dynamo states. {\it to be published in Fluid Dynamics
  and Dynamos in Astrophysics and Geophysics, eds. A.M. Soward,
C.A. Jones, D.W. Hughes, N.O. Weiss, Taylor \& Francis}
\bibitem[Chandrasekhar (1961)]{Chandra}
     \textsc{Chandrasekhar, S.} 1961
     {Hydrodynamic and Hydromagnetic Stability}, Oxford: 
     Clarendon Press.
\bibitem[Christensen \etal (1999)]{Christensenetal1999}
\textsc{Christensen, U., Olson, P., and Glatzmaier, G.A.} 1999
Numerical Modeling of the 
Geodynamo: A Systematic Parameter Study. {\it Geophys. J. Int.} {\bf 138} 
393--409.
\bibitem[Eschrich \& R\"udiger (1983)]{EschrichRudiger1983}
\textsc{Eschrich., K.-O., and R\"udiger, G.} 1983  A second-order correlation
approximation for thermal conductivity and Prandtl number of free
turbulence. {\it Astron. Nachr.} {\bf 304} 171--180.  
\bibitem[Glatzmaier \& Roberts (1995)]{GlatzmaierRoberts1995}
\textsc{Glatzmaier, G.A., and Roberts, P.H.} 1995 A three-dimensional convection
dynamo solution with rotating and finitely conducting inner core and
mantle. {\it Phys. Earth Plan. Inter.} {\bf 91} 63--75. 
\bibitem[Grote \& Busse (2000)]{GroteBusse2000}
\textsc{Grote, E., and Busse, F.H.} 2000 Hemispherical dynamos
generated by convection in rotating spherical shells. {\it
  Phys. Rev. E} {\bf 62} 4457--4460.
\bibitem[Grote \& Busse (2001)]{GroteBusse2001}
\textsc{Grote, E., and Busse, F.H.} 2001 Dynamics of convection and
dynamos in rotating spherical fluid shells. {\it Fluid Dyn. Res.} {\bf
  28} 349--368.   
\bibitem[Grote \etal (2001)]{Groteetal2001}
\textsc{Grote, E., Busse, F.H., and Simitev R.} 2001 {Buoyancy driven
  convection in rotating spherical shells and its dynamo action}, pp. 12-34
{\it in "High Performance Computing in Science and
  Engineering '01", E. Krause, W. J{\"a}ger, (eds.),  Springer}  
\bibitem[Grote \etal (1999)]{Groteetal1999}
\textsc{Grote, E., Busse, F.H., and Tilgner, A.} 1999
Convection-driven quadrupolar dynamos in rotating spherical shells,
{\it Phys. Rev. E} {\bf 60} R5025--R5028.
\bibitem[Grote \etal (2000)]{Groteetal2000}
\textsc{Grote, E., Busse, F.H., \& Tilgner, A.} 2000 Regular and chaotic spherical
dynamos. {\it Phys. Earth Planet. Inter.} {\bf 117} 259--272. 
\bibitem[Ishihara \& Kida (2002)]{IshiharaKida2002}
\textsc{Ishihara, N., and Kida, S.} 2002 Dynamo mechanism in a rotating spherical
shell:  competition between magnetic field and convection
vortices. {\it  J. Fluid Mech.} {\bf465} 1--32. 
\bibitem[Kageyama \& Sato (1997)]{KageyamaSato1997}
\textsc{Kageyama, A., and Sato, T.} 1997 Generation mechanism of a
dipole field by a magnetohydrodynamic dynamo. {\it Phys. Rev. E} {\bf
  55} 4617--4626. 
\bibitem[Katayama \etal (1999)]{Katayamaetal1999}
\textsc{Katayama, J.S., Matsushima, M., and Honkura, Y.} 1999 Some characteristics
of magnetic field behavior in a model of MHD dynamo thermally driven
in a rotating spherical shell. {\it Phys. Earth Planet. Inter.} {\bf
  111} 141--159.
\bibitem[Kida & Kitauchi (1998)]{KidaKitauchi1998}
\textsc{Kida, S., and Kitauchi, H.} 1998 Thermally driven MHD dynamo in a
rotating spherical shell. {\it Prog. Theor. Phys. Suppl.} {\bf 130} 121--136.
\bibitem[Kutzner \& Christensen (2002)]{KutznerChristensen2002} 
\textsc{Kutzner, C., and Christensen, U.R.} 2002 From stable dipolar towards
reversing numerical dynamos. {\it Phys. Earth Planet. Inter.} 
{\bf  131} 29--45.
\bibitem[Kutzner \& Christensen (2000)]{KutznerChristensen2000} 
\textsc{Kutzner, C., and Christensen, U.R.} 2000 Effects of driving
mechanisms in geodynamo models. {\it Geophys. Res. Lett.} {\bf
27} 29--32. 
\bibitem[Olson \etal 1999]{Olsonetal1999}
\textsc{Olson, P., Christensen, U., and Glatzmaier, G.A.} 1999
Numerical modeling of the geodynamo: Mechanism of field generation and
equilibration. {\it J. Geophys. Res.}, {\bf 104}, 10383--10404.
\bibitem[Simitev \& Busse (2003)]{SimitevBusse2003}
\textsc{Simitev, R. and Busse, F.H.} 2003 Patterns of convection in
rotating spherical shells. {\it New J. Phys.} {\bf 5} {97.1--97.20}.
\bibitem[Takahashi (2003)]{Takahashi2003}
\textsc{Takahashi, F., Matsushima, M., and Honkura,
Y.} 2003 Dynamo action and its temporal variation inside the tangent
cylinder in MHD dynamo simulations. {\it Phys. Earth Planet. Inter.}
{\bf 140} {53--71}. 
\bibitem[Tilgner \& Busse (1997)]{TilgnerBusse1997}
\textsc{Tilgner, A., and Busse, F.H.} 1997 Finite amplitude convection in rotating 
spherical fluid shells. {\it J. Fluid Mech.} {\bf 332} 359--376.
\bibitem[Wicht (2002)]{Wicht2002}
\textsc{Wicht J.} 2002 Inner-core conductivity in numerical dynamo simulations. {\it
  Phys. Earth Planet. Inter.} {\bf 132} 281--302.
\bibitem[Zhang (1994)]{Zhang1994}
\textsc{Zhang, K.} 1994 On coupling between the Poincar\'e equation and the heat
equation. {\it J. Fluid Mech.} {\bf 268} 211--229.
\bibitem[Zhang (1995)]{Zhang1995}
\textsc{Zhang, K.} 1995 On coupling between the Poincar\'e equation and the heat
equation: no-slip boundary condition. {\it J. Fluid Mech.} {\bf 284}
239--256.
\bibitem[Zhang \& Busse (1987)]{ZhangBusse1987}
\textsc{Zhang, K. and Busse, F.H.} 1987 On the onset of convection in rotating
spherical shells. {\it Geophys. Astrophys. Fluid Dyn.} {\bf 39}
119--147.
\bibitem[Zhang \& Busse (1989)]{ZhangBusse1989}
\textsc{Zhang, K. and Busse, F.H.} 1989 Convection driven magnetohydrodynamic
dynamos  in rotating spherical shells. {\it Geophys. Astrophys. Fluid
  Dyn.} {\bf 49} 97--116.
\bibitem[Zhang \& Busse (1990)]{ZhangBusse1990}
\textsc{Zhang, K., and Busse, F.H.} 1990  Generation of magnetic field by
convection in a rotating spherical fluid shell of infinite Prandtl
number. {\it Phys. Earth Planet. Inter.} {\bf 59} 208--222.
\end{thebibliography}
\end{document}